\documentclass[12pt]{article} 

\usepackage{latexsym} 
\usepackage{amssymb}  
\usepackage{epsfig}       

\newcommand{\beq}{\begin{equation}}
\newcommand{\eeq}{\end{equation}}
\newcommand{\ba}{\begin{array}}
\newcommand{\bea}{\begin{eqnarray}}
\newcommand{\ea}{\end{array}}
\newcommand{\eea}{\end{eqnarray}}

\newcommand\comment[1]{ \hbox{[{\it Comment suppressed here.}\/]} }
\newcommand\hide[1]{}

\renewcommand{\O}{ {\cal O} }

\newcommand{\Tr}{\hbox{Tr}}

\newcommand{\bx}{{\bf x}}
\newcommand{\by}{{\bf y}}
\newcommand{\bp}{{\bf p}}
\newcommand{\bq}{{\bf q}}
\newcommand{\bk}{{\bf k}}

\newcommand{\skipover}[1]{}

\newcommand{\C}{{\cal C}}

\makeatletter 

\def\appendix{\par                              
    \setcounter{section}{0}                     
    \setcounter{subsection}{0}
    \renewcommand{\theequation}{\Alph{section}.\arabic{equation}}
    \renewcommand{\thesection}{Appendix \Alph{section}
                \setcounter{equation}{0}  } 
}

\def\applabel#1{\@bsphack
  \protected@write\@auxout{}%
         {\string\newlabel{#1}{{\Alph{section}}{\thepage}}}%
  \@esphack}

\def\section{
\setcounter{equation}{0}        
\@startsection {section}{1}{\z@}{-3.5ex plus -1ex minus 
 -.2ex}{2.3ex plus .2ex}{\large\bf}}
\renewcommand{\theequation}{\arabic{section}.\arabic{equation}}

\def\subsection{\@startsection{subsection}{2}{\z@}{-3.25ex plus -1ex minus 
 -.2ex}{1.5ex plus .2ex}{\normalsize\bf}}

\def\subsubsection{\@startsection{subsubsection}{3}{\z@}{-3.25ex plus
 -1ex minus -.2ex}{1.5ex plus .2ex}{\normalsize}}

\makeatother   

\newsavebox{\eqlabel}

\makeatletter  
\newlength{\numblen}
\newsavebox{\eqnumb}
\def\@eqnnum{\savebox{\eqnumb}{\rm (\theequation)}%
\settowidth{\numblen}{\usebox{\eqnumb}}%
\makebox[\numblen][l]{\usebox{\eqnumb}~~~\usebox{\eqlabel}}}
\makeatother   

\newenvironment{equationwithlabel}[1]{ %
  \begin{equation}\label{#1} }{\end{equation}} 
\newcommand{\beql}[1]{\begin{equationwithlabel}{#1}}
\newcommand{\eeql}{\end{equationwithlabel}}

\begin{document}

\title{\bf Controlled nonperturbative dynamics 
of quantum fields out of equilibrium\\[2.ex]}

\author{J\"urgen Berges\footnote{Email: j.berges@thphys.uni-heidelberg.de} 
\\[1.ex]
{\normalsize Universit{\"a}t Heidelberg, 
Institut f{\"u}r Theoretische Physik}\\
{\normalsize Philosophenweg 16, 69120 Heidelberg, Germany}\\
}


\date{}

\begin{titlepage}
\maketitle
\def\thepage{}          

\vspace*{-0.5cm}
\begin{abstract}
We compute the nonequilibrium real-time
evolution of an \mbox{$O(N)$--symmetric} scalar quantum field theory 
from a systematic \mbox{$1/N$ expansion} of the $2PI$ effective action to 
next-to-leading order, which includes scattering and memory effects. 
In contrast to the standard $1/N$ expansion of the $1PI$ effective action,  
the next-to-leading order expansion in presence of  
a possible expectation value for the composite operator leads to a 
bounded time evolution
where the truncation error may be controlled by higher powers in $1/N$.
We present a detailed comparison with the leading-order results and 
determine the range of validity of standard mean field type approximations.    

We investigate ``quench'' and ``tsunami'' initial conditions
frequently used to mimic idealized far-from-equilibrium pion dynamics
in the context of heavy-ion collisions. For spatially homogeneous 
initial conditions we find three generic regimes, characterized by 
an early-time exponential damping, a parametrically slow (power-law) behavior 
at intermediate times, and a late-time 
exponential approach to thermal equilibrium. The different time scales 
are obtained from a numerical solution of the time-reversal invariant 
equations in $1\! +\! 1$ dimensions without further approximations. We 
discuss in detail the out-of-equilibrium behavior of the nontrivial 
\mbox{$n$-point} correlation functions as well as the evolution of a 
particle number distribution and inverse slope parameter.    

\end{abstract}

\end{titlepage}

\renewcommand{\thepage}{\arabic{page}}


\section{Introduction and overview}
\label{soverview}

Current and upcoming heavy ion collision experiments at RHIC
and the LHC have been an important motivation for the study of
quantum field theory in and out of equilibrium. With these
experiments one aims for a Quark Gluon Plasma produced in a 
transient state, finally releasing high multiplicities of
particles of which the lightest hadrons are the pions. One major
open question in the description of the underlying
physics concerns the justification of current quantum field theoretical
predictions based on equilibrium thermodynamics, local equilibrium or
linear response assumptions. Immediately related to this
appears the question for possible new signatures from  
processes which are far away from equilibrium. In contrast to a 
thermally equilibrated state, which
keeps no information about the details of the initial conditions,
nonequilibrium phenomena may help to understand in particular 
the earlier stages of a collision. 

A successful description of the dynamics of quantum fields away
from equilibrium is tightly related to the basic problem of how 
macroscopic ``dissipative'' behavior arises from time-reversal invariant
quantum dynamics. This is a fundamental question with most
diverse applications. The techniques, which have to be developed 
to understand the physics of heavy-ion collision experiments,
will be of relevance to the inflationary dynamics in the early 
universe, to the physics of  baryogenesis or even to nonequilibrium 
aspects of mesoscopic quantum devices.     
    
The description of nonequilibrium
quantum field theory from ``first principles'' can be based on
a path integral formulation using 
the Schwinger-Keldysh techniques \cite{Schwinger:1961qe}.   
In analogy to vacuum or thermal 
equilibrium quantum field theory one can construct a generating
functional for nonequilibrium Green's functions which contains all
quantum and statistical fluctuations for given initial
correlations or density matrix. 

However, controlled computational methods 
for the approximative solution of time evolution problems 
which respect the symmetries of the underlying theory are 
limited so far. In order to understand how macroscopic, effectively
irreversible behavior arises one cannot 
use phenomenological approaches which typically break the time-reflection
symmetry of the underlying theory.  
Simple (finite-loop) perturbative
descriptions can be unbounded and are known to break down at late 
times \cite{Jeon}.
While standard nonperturbative methods based on lattice Monte Carlo 
simulations have been successfully applied to thermal quantum field
theory in Euclidean time, a non-positive definite 
probability measure at real times already prevents their use.
In contrast, classical field theory can be simulated. One only
has to solve the classical field equations of motion sampled
with the appropriate initial condition. Of course, the latter is
not capable of treating genuine quantum effects but it should be
reliable when the number of field quanta in each mode is sufficiently
large. 
Another approach to which much work has been devoted restricts
the discussion to systems close to thermal equilibrium or to
effective descriptions based on a separation of scales in the 
weak coupling limit. For a recent review of the very interesting
developments in this field see Ref. \cite{Bodeker:2001pa}. 

Away from equilibrium, time-reflection invariant
approximations to quantum field theories have been almost 
uniquely based on mean field type
approximations (so-called 
``collisionless'' Hartree (-Fock) or leading order large-$N$ approximations) 
\cite{LOapp}.
It is a known problem that the latter approximations 
are unable to describe thermalization at late times\footnote{For 
improved results using inhomogeneous mean fields see Ref.\ \cite{LOinh}.}, 
and their extensive use
is based on the assumption that scattering effects 
may not change the picture dramatically for sufficiently early times. 
Efforts to go beyond mean field include a $1/N$ expansion of the 
generating functional for one-particle irreducible ($1PI$)
Green's functions beyond leading order \cite{LObeyond,LObeyondOsc}. 
However, the standard $1/N$ expansion of the $1PI$ effective action turns out
to be unbounded in time \cite{LObeyondOsc}. For time evolution problems it  
therefore fails to give a controlled expansion.

Recent progress came from a rather old concept based on
a loop expansion of the generating functional for two-particle 
irreducible ($2PI$) Green's functions as introduced in Refs.\
\cite{Cornwall:1974vz,Chou:1985es,Calzetta:1988cq,Phider}. 
It has been demonstrated in Ref.\ \cite{Berges:2000ur} 
for a scalar quantum field theory that a bounded time-evolution can 
be achieved without further approximations
from a three-loop expansion of the $2PI$ effective action. 
The three-loop approximation reflects all symmetries of the 
underlying theory and goes beyond the mean field level, including
scattering and memory effects. For very different nonequilibrium
initial conditions it was shown from a numerical solution in $1+1$ 
dimensions that the three-loop approximation leads to a universal
asymptotic behavior of quantum fields. The correlation
functions at late time are uniquely determined 
by the initial (conserved) energy density and approach the
equilibrium correlations of the corresponding three-loop thermal
field \mbox{theory \cite{Berges:2000ur}}.

At three-loop order the $2PI$ effective action comprises the classical
Boltzmann equation 
\cite{KadanoffBaym,Danielewicz,Calzetta:1988cq,Mrowczynski:1990bu,
Mrowczynski:1994hq,Blaizot:2001nr}.  
Without additional approximations the three-loop equations 
may be viewed as a ``quantum Boltzmann equation'' including
off-shell effects and resumming an infinite order of derivatives
\cite{AB1}, which are
difficult to include in kinetic descriptions  
\cite{Ivanov:2000tj,Joyce:2000uf}.
However, despite the fact that the loop expansion reveals a consistent
picture of the early- and late-time physics of quantum fields
without further assumptions, in absence of 
a small expansion parameter it is not a controlled approximation. 
In this work we will discuss a 
controlled nonperturbative approximation.\\

Our aim is to study a quantum field theory which captures 
aspects of low energy pion physics and which is simple enough that 
one can perform a quantitative treatment.
The basic ingredient for the effective low energy description
is the approximate chiral symmetry of the classical 
action for quantum chromodynamics (QCD). 
In QCD for $N_f$ flavors the chiral $SU_L(N_f)\times SU_R(N_f)$
is broken to a vector-like $SU(N_f)$, where the pions appear
as the corresponding Goldstone bosons. For two flavors
the $SU(2) \times SU(2)$ is locally isomorphic to 
an $O(4)$ symmetry which we will consider in the following. 
Unlike in chiral perturbation theory we
will study here the corresponding {\it linear} model 
(``linear sigma model'') for
the light scalar and pseudo-scalar degrees of freedom 
which constitute the $O(N=4)$--vector 
($\sigma$, $\vec{\pi}$). For various $N$ this model has been
extensively studied in thermal equilibrium, or nonequilibrium
as in Refs.\ \cite{LOapp,LObeyond}\footnote{For $N=1$ it also describes 
the long-range fluctuations of the QCD
critical point \cite{Berges:1999rc}.}.

We employ a systematic {\bf 1/N expansion
of the O(N)--symmetric 2PI effective action} $\Gamma[\phi,G]$, which
does not only depend on a possible expectation value of the
quantum field, $\phi$, but also on an expectation value 
of the time-ordered field bilinear, $G$ 
(cf.\ Sects.\ \ref{section2pi}, \ref{Nexpansion}). 
Though $N=4$ may be used already as an expansion parameter, we will 
frequently consider larger values of $N$ for the sake of
a quantitative treatment. We compute the $2PI$ effective action
to next-to-leading order in $1/N$, which goes beyond the mean field-type
leading-order approximation and includes
scattering and memory effects. This approximation leads to a bounded
time evolution for which the truncation error may be controlled
by higher powers in $1/N$. In this case the employed expansion 
provides a {\bf controlled
nonperturbative approximation} which is not restricted to the
weak coupling regime. Time-reflection invariance and 
conservation of energy is preserved at each order in the
$1/N$ expansion.
\begin{figure}[t]
\begin{center}
\epsfig{file=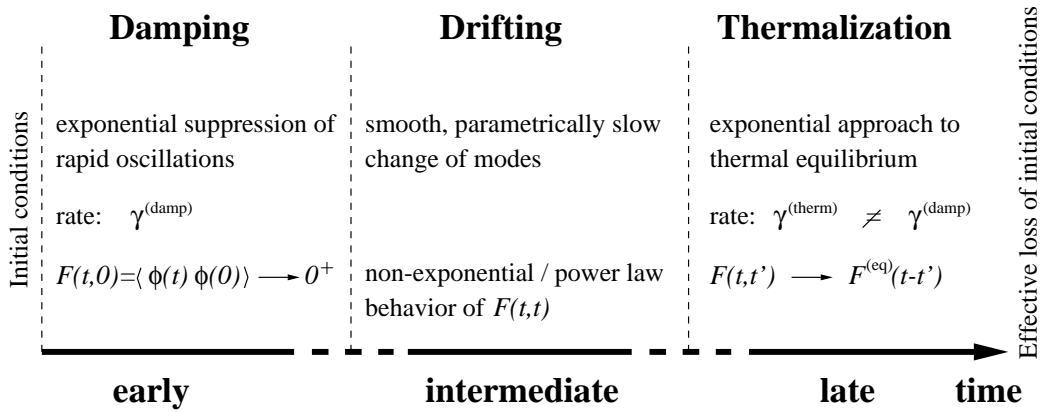,width=5.5cm,angle=-90}
\end{center}
\vspace*{-0.5cm}
\caption{For the $O(N)$--symmetric scalar theory in $1\! +\! 1$
dimensions we find three generic time regimes: An approximately
exponential damping of oscillations characteristic for early times 
with inverse rate 
$\tau^{\rm (damp)} \sim 1/\gamma^{\rm (damp)}$, a parametrically slow 
or power law drifting at intermediate times, and a late-time exponential 
approach to thermal equilibrium
with $\tau^{\rm (therm)} \sim 1/\gamma^{\rm (therm)}$.
Both for the employed ``quench'' and the ``tsunami'' initial condition
$\tau^{\rm (damp)}\,\, \ll\,\, \tau^{\rm (therm)}$. None of the
described different time scales are captured in the standard 
leading-order or mean field type approximations. The inclusion
of scattering effects is crucial and the leading-order approximation
is approximately valid only for $t \ll 1/\gamma^{\rm (damp)}$.}
\label{Figoverview}
\end{figure}
For a discussion of similar approximation schemes applied to    
an $N$-component quantum anharmonic 
oscillator see also \mbox{Ref.\ \cite{Resummed}}. 
For $N\! =\! 1$ the current approximation includes the type of
diagrams involved in the three-loop approximation 
of the $2PI$ effective action used in Refs.\ 
\cite{Berges:2000ur,AB1}. The precise correspondence 
can be found by neglecting all contributions of the current approximation
beyond three-loop and rescaling the four-point 
function ``$L \to L/3$'' in the three-loop approximation 
of Ref.\ \cite{Berges:2000ur}. 
 
We study the time evolution for two initial condition scenarios away
from equilibrium. The first one corresponds to 
a ``quench'' (Sect.\ \ref{ssQuench}) often employed to mimic 
the situation of a 
rapidly expanding hot initial state which cools on time scales much 
smaller than the relaxation time of the fields \cite{Rajagopal:1993ah}. 
Initially at high temperature we consider the relaxation processes 
following an instant ``cooling'' described by a sudden drop in the 
effective mass.
The second scenario is characterized by initially densely populated modes in
a narrow momentum range around $\pm \bp_{\rm ts}$ 
(cf.\ \mbox{Sect.\ \ref{sstsun}}).
This initial condition has been termed ``tsunami'' in 
\mbox{Ref.\ \cite{Tsunami}}
and is reminiscent of colliding wave packets 
moving with opposite and equal momentum. A similar nonthermal and radially
symmetric distribution of highly populated modes may also be encountered in a 
``color glass condensate'' at saturated gluon density with typical
momentum scale $p_{\rm ts}$ \cite{McLerran:1994ni,RDP}. 
Of course,
a sudden change in the two-point function of a previously equilibrated system
or a peaked initial particle number distribution are general enough 
to exhibit characteristic properties
of nonequilibrium dynamics for a large variety
of physical situations.  
  
To reduce the numerical ``cost'' of
following the evolution to sufficiently large times
we present results for the $1+1$ dimensional
quantum field theory with spatially homogeneous correlations.
We stress that in $1+1$ dimensions many realistic
questions can not be addressed, in particular since there is no 
spontaneous symmetry breaking. The $1+1$ dimensional
dynamics may help to understand the symmetric regime relevant for 
sufficiently high energy densities. However, the current results should be
understood as a first controlled, quantitative approach to understand
the dynamics of quantum fields far away from equilibrium.   
 
Independent of the details of the initial conditions we find
the following characteristic time scales summarized in 
\mbox{Fig.\ \ref{Figoverview}}. The nontrivial two- and
four-point functions at next-to-leading order in $1/N$
oscillate with initial frequency proportional to the
renormalized initial mass $M_{\rm INIT}$. 
All correlation functions quickly approach an exponentially damped
behavior. A characteristic rate $\gamma^{\rm (damp)}_0$ can be obtained 
from the zero mode of the unequal-time two-point function
$F(t,0;p=0) = \langle \phi(t) \phi(0) \rangle_{p=0}$ with a 
corresponding time scale $\tau^{\rm (damp)}$ proportional to the inverse rate
(\mbox{Sect.\ \ref{ssEarlytime}}). 
In this time range correlations with the initial state are effectively
suppressed and asymptotically $F(t,0;p=0) \to 0^+$. 

Both for the ``quench'' and the
``tsunami'' we observe that after the characteristic damping time
the system is still far away from equilibrium. We find very different
rates for damping and for the late-time exponential approach to
thermal equilibrium\footnote{For the employed 
spatially homogeneous initial conditions we do not
observe a power-law ``tail'' for the late-time evolution (see also
Ref.\ \cite{Arnold:1998gh}).}, i.e.\
$\tau^{\rm (damp)}\,\, \ll\,\, \tau^{\rm (therm)}$.
Though the exponential damping at early times
is crucial for an effective loss of details of the initial conditions
--- a prerequisite for the approach to equilibrium --- 
it does not determine the time scale for thermalization.
At intermediate times the evolution of correlation
functions is found to be characterized by a parametrically slow 
change (drifting) for times typically much smaller 
than $\tau^{\rm (therm)}$. 
We observe characteristic power law behavior
$|F(t,t;p) - F(0,0;p)| \sim t^{{\rm (power)}_p}$ reminiscent
of a hydrodynamic regime 
(\mbox{Sect.\ \ref{sDrifting}}). The large-time limit 
of correlation functions is determined by the (conserved)
energy density and independent of the detailed initial conditions.

Apart from the three qualitatively different time regimes,
whose presence is rather insensitive to the details of the initial 
conditions, there are nonequilibrium phenomena specific to
a ``quench'' or a ``tsunami'' 
(cf.\ \mbox{Sect.\ \ref{Sectnonequilibriumdynamics}}). 
After quenching from an initial 
thermal distribution with temperature $T_0$ we find
that the system relaxes asymptotically to an effective particle 
number distribution with temperature (inverse slope parameter) 
$T < T_0$. During the nonequilibrium evolution 
the nontrivial two- and four-point 
correlation modes show no substantially different behavior
for low or for high momenta in absence of spontaneous symmetry
breaking. In contrast, the ``tsunami'' initial condition can
lead to strong differences in the dynamics for the low 
($p \ll p_{\rm ts}$) and the high lying modes ($p \gg p_{\rm ts}$).
In particular, for the low momentum modes 
the inverse slope parameter quickly shoots up in response 
to the ``tsunami'' and can reach several $T_{\rm EQ}$ at 
intermediate times. The high momentum modes show the opposite
effect and ``cool'' down slightly, before the inverse slope
$T(x^0;p)$ approaches its asymptotic constant value $T_{\rm EQ}$
(cf.\ Sect.\ \ref{ssInverseslope}).

New phenomena of far-from-equilibrium physics include also the
possible appearance of strongly interacting low momentum modes
at intermediate times. For the employed ``tsunami'' of
\mbox{Sect.\ \ref{sstsun}} the effective four-point function
$\lambda_{\rm eff}(x^0;p=0)/6N$ reaches several times
its initial value before it relaxes to its asymptotic
late-time result, which is smaller that the initial value $\lambda/6N$
(cf.\ \mbox{Sect.\ \ref{ssStronginteractions}}). 
The strong renormalization
of the effective four-point function at intermediate times also 
stresses the fact that effective descriptions based on a clear 
separation of scales can be difficult to achieve for far-from-equilibrium
processes.

Sects.\ \ref{section2pi} and \ref{Sectnonequilibrium} present
a short introduction to the generating functional for
nonequilibrium $2PI$ Green's functions. In Sects.\ \ref{Nexpansion}
and \ref{Secteom} we compute the $2PI$ effective action to 
next-to-leading order in $1/N$ and derive the equations of motion.
\mbox{Sect.\  \ref{Sectnumimp}} discusses the numerical implementation.
The numerical results are given in
\mbox{Sect.\ \ref{Sectnonequilibriumdynamics}}, where we also present 
a comparison with leading-order results and 
determine the range of validity of mean field type 
approximations. We end with a conclusion and an outlook in 
\mbox{Sect.\ \ref{SectCAO}}.

\section{$2PI$ effective action}
\label{section2pi}

We consider a quantum field theory with classical action 
$S = \int {d}^{d+1}x\, {\cal L}$,
\beq
\label{classical}
{\cal L}(x;\varphi) = \frac{1}{2} 
\partial_{x^0} \varphi_a
\partial_{x^0} \varphi_a
-\frac{1}{2} \partial_{\bx} \varphi_a
\partial_{\bx} \varphi_a
- \frac{1}{2} m^2 \varphi_a \varphi_a
- \frac{\lambda}{4! N} \left(\varphi_a\varphi_a\right)^2
\eeq
where $\varphi_a(x)$, $x\equiv (x^0,\bx)$, is a real, 
scalar field with $a=1,\ldots,N$ components
(summation over repeated indices is implied). All correlation functions 
of the quantum theory can be obtained from the effective 
action $\Gamma[\phi,G]$, i.e.\ the generating functional
for two-particle irreducible ($2PI$) Green's functions 
parametrized by the field $G_{ab}(x,y)$ representing an
expectation value of the time ordered composite
$T \varphi_a(x) \varphi_b(y)$ 
and the macroscopic field $\phi_a(x)$ given by the expectation 
value of $\varphi_a(x)$. 
A discussion of the defining functional integral of the
$2PI$ effective action can be found in Ref.\ \cite{Cornwall:1974vz}. 
Following
\cite{Cornwall:1974vz} it is convenient to parametrize the $2PI$ 
effective action as
\beq
\Gamma[\phi,G] = S[\phi] + \frac{i}{2} \Tr\ln G^{-1} 
          + \frac{i}{2} \Tr\, G_0^{-1} G
          + \Gamma_2[\phi,G] +{\rm const} 
\label{2PIaction}
\eeq 
which expresses $\Gamma$ in terms of the classical action $S$
and correction 
terms including the function $\Gamma_2$ which is discussed below. 
Here the classical inverse propagator
$i G_{0,ab}^{-1}(x,y;\phi)=
\delta^2 S[\phi]/\delta \phi_a(x) \delta \phi_b(y)$ 
is given by
\bea
i G^{-1}_{0,ab}(x,y;\phi) &=& - \left( \square_x + m^2 
+ \frac{\lambda}{6 N}\, \phi_c(x)\phi_c(x) \right) \delta_{ab}
\delta^{d+1}(x-y) \nonumber\\ 
&& - \frac{\lambda}{3 N}\, \phi_a(x) \phi_b(x) \delta^{d+1}(x-y) \, .
\eea
In absence of external sources
physical solutions require 
\bea
\frac{\delta \Gamma[\phi,G]}{\delta \phi_a(x)} &=& 0 \, , 
\label{phistationary}\\
\frac{\delta \Gamma[\phi,G]}{\delta G_{ab}(x,y)} &=& 0 \, .
\label{stationary}
\eea
Taking the derivative of (\ref{2PIaction}) with respect to $G$ one
observes that the second stationarity condition (\ref{stationary}) is
equivalent to the exact Schwinger--Dyson equation for the
propagator:
\beq
G_{ab}^{-1}(x,y) = G_{0,ab}^{-1}(x,y;\phi) - \Sigma_{ab}(x,y;\phi,G)
\label{SchwingerDyson}
\eeq
with the proper self energy   
\beq
\Sigma_{ab}(x,y;\phi,G) =  
2 i\, \frac{\delta \Gamma_2[\phi,G]}{\delta G_{ab}(x,y)} \, .
\label{exactsigma}
\eeq
The simple relation between the self energy     
and the correction term $\Gamma_2$ motivates the 
representation (\ref{2PIaction}) of the $2PI$ effective action. 
It is instructive to observe from 
(\ref{exactsigma}) that $\Gamma_2$ must be two--particle 
irreducible (cf.\ also the detailed discussion in 
Ref.\ \cite{Cornwall:1974vz}). 
For (\ref{SchwingerDyson}) to be an 
identity $\Sigma$ is to be interpreted as the 
proper self energy where only one--particle 
irreducible graphs can contribute. Suppose $\Gamma_2$ has
a two--particle reducible contribution of the form
$\tilde{\Gamma} G G \tilde{\Gamma}'$.
Then $\Sigma$ has a contribution of the form 
$\tilde{\Gamma} G \tilde{\Gamma}'$ since it is given by a derivative
of $\Gamma_2$ with respect to $G$. Such a structure
cannot occur for the proper self energy and two--particle 
reducible contributions to $\Gamma_2$ must be absent. 

Truncated Schwinger--Dyson equations are often used to
describe nonperturbative physics.  
Typically the employed approximation is based on some
ansatz for $\Sigma$ which in turn corresponds to an efficient 
resummation of a large number of graphs. It is 
apparent from the simple relation (\ref{exactsigma}) that      
the $2PI$ effective action may be used as a powerful tool  
to obtain resummation schemes in a systematic way. 

The contribution $\Gamma_2$ in (\ref{2PIaction}) is given by all 
$2PI$ graphs with the propagator lines set equal to $G$ 
\cite{Cornwall:1974vz}.
(A graph is two--particle irreducible if it does not
become disconnected upon opening two propagator lines.) 
In presence of a nonvanishing expectation value $\phi$ the
$2PI$ graphs are constructed from two kinds of vertices
described by the effective \mbox{interaction Lagrangian}
\beq
{\cal L}_{\rm INT}(x;\phi,\varphi) =  
- \frac{\lambda}{6 N} \phi_a(x)\varphi_a(x)\varphi_b(x)\varphi_b(x)
- \frac{\lambda}{4! N} \Big(\varphi_a(x)\varphi_a(x)\Big)^2 \, .
\eeq
The effective interaction can be obtained from (\ref{classical}) 
by shifting the field $\varphi \rightarrow \varphi + \phi$.
The terms cubic and higher in $\varphi$ define the vertices
from which the $2PI$ graphs contributing to $\Gamma_2$ can be
constructed. 

The relation of the $2PI$ effective action to the conventional generating
functional for {\it one--}particle irreducible Green's functions is 
very simple. The $1PI$ effective action corresponds to $\Gamma[\phi,G]$ at 
that value of $G$ for which (\ref{stationary}) holds, i.e.\
\beq
\Gamma[\phi] = \Gamma[\phi,G_{\rm stat}] \quad , \qquad
{\frac{\delta \Gamma[\phi,G]}{\delta G(x,y)}}\Big|_{G=G_{\rm stat}}=0 \, .
\eeq

\section{Nonequilibrium Green's functions}
\label{Sectnonequilibrium}

We will use the $2PI$ effective action to study the time evolution
of \mbox{$n$-point} functions for given initial correlations at time $t_I$.
The initial $n$--point functions 
$\Tr\{ \bar{\rho}(t_I) \varphi(t_I,\bx_1) \ldots \varphi(t_I,\bx_n)\}$ 
may include a mixed--state density matrix $\bar{\rho}(t_I)$
for which $\Tr\{\bar{\rho}^2(t_I)\} < 1$. This is necessary to allow for
general nonequilibrium initial conditions and, in particular,  
if the initial density matrix describes thermal equilibrium. 
Often the initial conditions of an experiment --- 
one may think of a scattering 
experiment of particles at an initial time $t_I$ far before they reach the 
interaction region --- 
may be described by only a few lowest $n$--point functions. For a 
setup which can be described in terms of an initial Gaussian density matrix
the nonvanishing initial correlations are completely specified by $\phi$ and
$G$ and their first order time derivatives\footnote{
For a Gaussian initial mixed-state density matrix there are three 
independent variances which can be chosen as $G$, $H$ and $K$ 
defined in (\ref{inG}). Note that
the initial correlation $\Tr\{ \bar{\rho}(t_I) \varphi(t_I,\bx)
\partial_{y^0}\varphi(y^0,\by)\}|_{y^0=t_I}$ is not independent 
because of the commutation 
relation between $\varphi$ and $\partial_{y^0}{\varphi}$. 
A detailed discussion of Gaussian initial
conditions can be found in \mbox{Ref.\ \cite{CHKM}}. 
} at $t_I$, i.e.\
\bea
\phi(t_I,\bx) &=& \Tr\{ \bar{\rho}(t_I) \varphi(t_I,\bx)\}\, ,
\label{inphi}
\\
\dot{\phi}(t_I,\bx)
&=& \Tr\{ \bar{\rho}(t_I) \partial_{x^0} \varphi(x^0,\bx)\}|_{x^0=t_I}\, ,
\nonumber
\eea
and the initial two-point functions
\bea
G(t_I,\bx;t_I,\by) &=& \Tr\{ \bar{\rho}(t_I) \varphi(t_I,\bx)
\varphi(t_I,\by)\} - \phi(t_I,\bx)\phi(t_I,\by) , 
\label{inG}
\\
H(t_I,\bx;t_I,\by)
&=& \Tr\{ \bar{\rho}(t_I) \partial_{x^0} \varphi(x^0,\bx)
\varphi(t_I,\by)\}|_{x^0=t_I} - \dot{\phi}(t_I,\bx)\phi(t_I,\by)\, ,
\nonumber\\
K(t_I,\bx;t_I,\by) 
&=& \Tr\{ \bar{\rho}(t_I) \partial_{x^0} \varphi(x^0,\bx)
\partial_{y^0} \varphi(y^0,\by)\}|_{x^0=y^0=t_I} 
- \dot{\phi}(t_I,\bx)\dot{\phi}(t_I,\by)\nonumber .
\eea 
We will assume in the following that the initial conditions
can be described in terms of a Gaussian density matrix.
It should be stressed at this point that a specific initial 
condition only restricts the experimental setup and represents
no approximation for the time evolution. For 
$t>t_I$ irreducible higher \mbox{$n$--point} functions 
are induced by the time evolution of the interacting theory.
The full $2PI$ effective action $\Gamma[\phi,G]$
contains all possible information about $n$-point functions for
given initial conditions. 
 
A more general $\bar{\rho}(t_I)$ involves the specification of nonzero 
higher initial \mbox{$n$--point} functions than (\ref{inphi}) and
(\ref{inG}). The $2PI$ effective action in presence
of a more general initial density matrix is discussed in 
\cite{Calzetta:1988cq,Berges:2000ur}. 
In general, a non-Gaussian $\bar{\rho}(t_I)$
leads to additional source terms in the
generating functional for $2PI$ Green's functions and $\Gamma$
is no longer of the simple form (\ref{2PIaction}).
These source terms are non-vanishing at  
time $t_I$ only and allow for the specification of the
additional nonzero initial correlations.     

For Gaussian initial conditions
the time evolution of $n$-point functions 
$\Tr\{ \bar{\rho}(t_I) \varphi(x) \ldots \varphi(y)\}$ 
is completely specified by (\ref{inphi}) and (\ref{inG}) and known
$\Gamma[\phi,G]$. It should be noted that the $n$-point functions
of the initial time problem differ from those used to
compute $S$--matrix elements in conventional scattering
theory. The $n$-point functions for the initial time problem
involve the trace over products of field operators. Stated
differently, one computes matrix elements between same
states. These are sometimes called ``$in-in$''
matrix elements in contrast to ``$in-out$'' matrix 
elements used in conventional scattering theory.
The construction of the generating 
functionals for the respective $n$-point functions  
is identical up to the time path. The trace over products of 
field operators implies that the generating
functional used for initial time problems 
involves a closed time path $\C$ \cite{Schwinger:1961qe}.
The time path starts at the initial time $t_I$ and runs 
along the (real) time axis to some final time $t$ and back to $t_I$. 
The largest time of the path $t$ is kept as a parameter and is evolved
in the time evolution equations for the fields as described
in Sect.\ \ref{ssInitialvalue}.\footnote{ 
Formulations of closed time path generating
functionals often exhibit an infinite time interval $]- \infty,\infty[$. 
In this case
the number of field labels has to be doubled to distinguish the fields
on the underlying closed contour. Causality implies that for any 
$n$-point function at finite times $t_i$ $(i=1,\ldots,n)$ the 
contributions of the infinite time path employed in these formulations 
cancel for times larger than ${\rm max}(t_i)$. For our purposes of a
possible numerical implementation as an initial-value problem the infinite 
time path is not useful.}

\section{$1/N$ expansion of the $2PI$ effective action}
\label{Nexpansion} 

In the interacting theory the $2PI$ effective action cannot be 
computed exactly. In this work we employ 
a $1/N$ expansion of the $2PI$ effective action $\Gamma[\phi,G]$ where $N$ 
is the number of field components. 
The leading order approximation is found by
considering only those $2PI$ diagrams where the summation over internal lines 
yields a factor greater than or equal to $N^l$ with $l$ the
number of loops in the diagram. The next-to-leading order diagrams
yield a factor $N^{l-1}$, next-to-next-to-leading order $N^{l-2}$
and so forth. 
This defines a systematic  
classification in terms of powers of $1/N$ for $2PI$ diagrams 
contributing to $\Gamma[\phi,G]$. For a discussion of similar 
approximation schemes applied to quantum mechanics see also 
Ref.\ \cite{Resummed}. 
In the large-$N$ limit $\phi_a$ can be taken to be ${\cal O} (\sqrt{N})$, 
hence $G_{ab}$ is ${\cal O} (1)$ \cite{Cornwall:1974vz}. The leading order 
term is then proportional to $N$ and the next-to-leading order term
is proportional to one, as will be considered in the following.

\subsection{$2PI$ effective action to next-to-leading order}

\begin{figure}[t]
\begin{center}
\epsfig{file=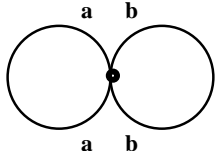,width=2.8cm}
\end{center}
\vspace*{-0.6cm}
\caption{$LO$ contribution to the $2PI$ effective action.} 
\label{LOfig}
\end{figure}

For simplicity we will concentrate in the following on the
symmetric regime where it is sufficient to consider 
$\Gamma[\phi=0,G] \equiv \Gamma[G]$. We write  
\beq
\Gamma_2[G]= \Gamma_2^{\rm LO}[G] 
          + \Gamma_2^{\rm NLO}[G] 
          + \ldots
\eeq
where $\Gamma_2^{\rm LO}$ denotes the leading order (LO) 
and $\Gamma_2^{\rm NLO}$ the next-to-leading order (NLO) contributions. 
The dots indicate terms beyond the NLO approximation. 
The LO contribution to $\Gamma_2[G]$ is given by
only one diagram shown in Fig.\ \ref{LOfig},
\beq
\Gamma_2^{\rm LO}[G] = - \frac{\lambda}{4! N} 
  \int_{\C} {d}^{d+1}x\, G_{aa}(x,x) G_{bb}(x,x) \, .  
\label{LOcont}
\eeq  

\begin{figure}[b]
\begin{center}
\epsfig{file=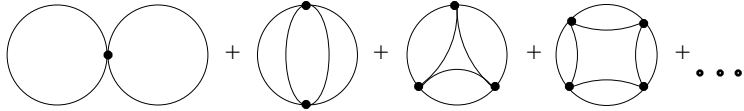,width=9.8cm}
\end{center}
\vspace*{-0.6cm}
\caption{$NLO$ contribution to the $2PI$ effective action.} 
\label{NLOfig}
\end{figure}

\begin{figure}[t]
\begin{center}
\epsfig{file=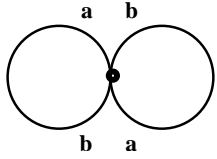,width=2.8cm}
\end{center}
\vspace*{-0.6cm}
\caption{$NLO$ ``double bubble'' contribution.} 
\label{NLObubblefig}
\end{figure}

\begin{figure}[b]
\begin{center}
\epsfig{file=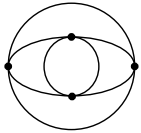,width=1.8cm}
\end{center}
\vspace*{-0.6cm}
\caption{The five--loop $2PI$ ``eye''-diagram contributes at $NNLO$.} 
\label{EYEfig}
\end{figure}

Omitting combinatorial factors Fig.\ \ref{NLOfig} indicates the diagrams 
contributing to $\Gamma_2[G]$ in next-to-leading order. 
The first
diagram of the series corresponds to the double bubble diagram 
shown in Fig.\ \ref{NLObubblefig}. 
It is followed by a three--loop diagram shown and
then each subsequent diagram of the NLO series can be constructed 
from the preceding one by introducing at one vertex an additional loop
as in Fig.\ \ref{NLOfig}. 
One may view these contributions as a three--loop
diagram with an effective four--vertex containing a ``chain'' of bubbles. 
The double bubble and the infinite series of
closed chain diagrams are the only graphs contributing to 
$\Gamma_2[G]$ to NLO. 
For instance, the other $2PI$ five--loop
diagram shown in Fig.\ \ref{EYEfig} only contributes to NNLO. 
It is straightforward to sum the infinite number of NLO diagrams 
analytically. One finds 
\beq
\Gamma_2^{\rm NLO}[G] =  \frac{i}{2} \int_{\C} {d}^{d+1}x\, 
\ln [\, {\bf B}(G)\, ] (x,x)
\label{NLOcont} 
\eeq
where
\beq
{\bf B}(x,y;G) = \delta_{\C}^{d+1}(x-y)
+ i \frac{\lambda}{6 N}\, G_{ab}(x,y)G_{ab}(x,y) \, .
\label{Feq}
\eeq
It is instructive to expand the RHS of (\ref{NLOcont}) to see the 
correspondence with the series of graphs in Fig.\ \ref{NLOfig}
\bea
\lefteqn{\int_{\C} {d}^{d+1}x\, \ln [\, {\bf B}(G)\, ] (x,x)
=  \int_{\C} {d}^{d+1}x\, 
\Big( i \frac{\lambda}{6 N}\, G_{ab}(x,x)G_{ab}(x,x) \Big)}
\nonumber\\[0.1cm]
&& -\frac{1}{2} \int_{\C} {d}^{d+1}x\,{d}^{d+1}y\,
\Big( i \frac{\lambda}{6 N}\, G_{ab}(x,y)G_{ab}(x,y) \Big)
\Big( i \frac{\lambda}{6 N}\, G_{a'b'}(y,x)G_{a'b'}(y,x) \Big)
\nonumber\\[0.2cm]
&& + \ldots
\eea
The contributions (\ref{LOcont}) and (\ref{NLOcont})
constitute the NLO approximation for $\Gamma[G]$. 
We will use this without further approximations to derive 
the time evolution equation for $G$ 
in Sect.\ \ref{Secteom}, which parametrizes 
the generating functional $\Gamma[G]$. Once $G$ is known
for all times of interest the theory is ``solved'' in the 
symmetric regime to NLO: All correlation functions can 
be obtained from derivatives of the generating functional
(\ref{2PIaction}) and are given by (summed) diagrammatic 
expressions in terms of $G$. 

\subsection{Comparison with conventional $1/N$ schemes}

We stress that in the above $2PI$ scheme the counting 
of orders of $1/N$ is done for diagrams with propagator lines
associated to the field $G$. The time evolution for the
propagator field is governed by a differential
equation in $G$ which is explicit in time and can be solved
by standard techniques for given initial conditions (cf.\ Sect.\ 
\ref{ssInitialvalue}). 
If not for a perturbative approximation there is 
no need to express $G$ in terms of the classical propagator
$G_0$. However, to compare with standard $1/N$ expansions 
for the $1PI$ effective action $G$ can be written as an
infinite series parametrized by $G_0$. 

To leading order there is no difference 
to the conventional $1/N$ expansion scheme \cite{Cornwall:1974vz}. 
The equivalence can be observed using 
(\ref{exactsigma}) and iteratively expanding the result for the
full propagator (\ref{SchwingerDyson}) in powers 
of the coupling $\lambda$. This yields the solution of 
$G$ in terms of an infinite series of diagrams 
with propagator lines associated to the classical propagator $G_0$.
Replacing this solution for $G$ in (\ref{LOcont}) one recovers  
the conventional ``bubble sum'' expression for the leading order
$1PI$ effective action. 

\begin{figure}[bt]
\begin{center}
\epsfig{file=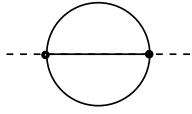,width=2.5cm}
\end{center}
\vspace*{-0.6cm}
\caption{``Setting sun'' type contribution to the self energy
with lines $G$.} 
\label{SELFfig}
\end{figure}

Beyond leading order the counting for 
the $2PI$ effective action differs from the conventional $1/N$ 
expansion. The two schemes only differ by higher order terms. 
As an example, one may consider the three--loop
diagram in Fig.\ \ref{NLOfig}. From (\ref{exactsigma}) one finds that
this diagram leads to the ``setting sun'' type contribution to the 
self energy shown in Fig.\ \ref{SELFfig}, where lines correspond to the full 
propagator $G$. Expressing the resulting $G$ in terms of $G_0$ 
one observes, in particular, the resummation of an infinite series of
``ladder'' diagrams where two lines repeatedly exchange momenta 
by one-loop subdiagrams connecting the two lines. This infinite series 
of diagrams is not included in the conventional $1/N$ expansion at
any finite order in $1/N$. A detailed discussion of the importance of 
these diagrams in the context of transport
coefficients can be found in Refs.\ \cite{Jeon,CH}.  

In Sect.\ \ref{Sectnonequilibriumdynamics} we will observe that the 
employed approximation for the $2PI$ effective action  
leads to a bounded time evolution for the employed initial conditions,
both at LO and NLO. In this case also the neglected term 
$\Gamma_2[G]-\Gamma_2^{\rm LO}[G]-\Gamma_2^{\rm NLO}[G]$
is finite at all times for a bounded full $\Gamma[G]$ or $\Gamma_2[G]$.
The truncation error may therefore be controlled by its suppression
with higher powers in $1/N$. 
This is in contrast to the conventional $1/N$ expansion
for the $1PI$ effective action where terms grow unboundedly
with time \cite{LObeyondOsc}. In the case of contributions
which grow proportional to powers of $t^a/N$ for some $a > 0$
such an expansion would break down at times of order $N^{1/a}$.

The $2PI$ effective action is known to lead to a non-secular time 
evolution already for the three-loop approximation of the $2PI$ effective
action as has been pointed out in Ref.\ \cite{Berges:2000ur}. 
In Ref.\ \cite{Resummed} the $N$-component anharmonic oscillator, 
resumming an equivalent set of diagrams as discussed here, has also 
been shown to exhibit a bounded behavior. So far, we have not found initial
conditions where this property could not be observed for the 
current approximation. To find the desirable analytical 
proof for arbitrary initial conditions is a difficult task in
the presence of scattering and memory effects involved in the 
NLO time evolution, as is explained in the following. 

\section{Equation of motion}
\label{Secteom}

\subsection{Schwinger--Dyson equation}
\label{ssSchwingerDysoneq}

According to (\ref{SchwingerDyson}) the exact inverse propagator  
can be written as 
\bea
i\, G_{ab}^{-1}(x,y) = - (\square_x + m^2)\, \delta_{ab} \,
\delta_{\C}^{d+1}(x-y) - i \Sigma_{ab}(x,y;G) 
\label{ginversexact}
\eea
where the full self energy $\Sigma$ is given by (\ref{exactsigma}).   
In next-to-leading order in $1/N$ the approximation for the self
energy follows from the derivative of (\ref{LOcont}), (\ref{NLOcont})
with respect to $G_{ab}$. One finds
\beq
\Sigma_{ab}(x,y;G) = 
-i  \frac{\lambda}{6 N}\, G_{cc}(x,x) \delta_{ab}\, \delta_{\C}^{d+1}(x-y)
-i \frac{\lambda}{3 N}\, G_{ab}(x,y) {\bf B}^{-1}(x,y;G)  
\label{sigmanext}
\eeq
with ${\bf B}$ given by (\ref{Feq}). Multiplying (\ref{Feq}) with
${\bf B}^{-1}$ and using the identity
$\int_{\C} {d}^{d+1}z\, {\bf B}(x,z) {\bf B}^{-1}(z,y)=
\delta_{\C}^{d+1}(x-y)$ one observes that the inverse of ${\bf B}$
obeys
\bea
{\bf B}^{-1}(x,y;G)&=&\delta_{\C}^{d+1}(x-y)-
i\, I(x,y) \, .
\label{Binverse}
\eea
Here the function $I(x,y)$ resums the ``chain'' of bubble graphs 
discussed in Sect.\ \ref{Nexpansion}, i.e.\
\bea
I(x,y)&=&\frac{\lambda}{6 N}\, G_{ab}(x,y) G_{ab}(x,y)
 -\, i\, \frac{\lambda}{6 N} \int_{\C} {d}^{d+1}z\, I(x,z)
G_{ab}(z,y) G_{ab}(z,y)  \, .
\nonumber\\
\label{ieqab}
\eea
Rewriting (\ref{Binverse}) and 
(\ref{ieqab}) we note that the combination 
\beq
\frac{\lambda}{6 N}\, {\bf B}^{-1}(x,y)
= \frac{\lambda}{6 N} \Big( \delta_{\cal C}^{d+1}(x-y) 
- i \frac{\lambda}{6 N}\, \int_{\cal C}\! d^{d+1} z\, 
{\bf B}^{-1}(x,z) G_{ab}(z,y) G_{ab}(z,y)\Big)  \nonumber
\label{fourvertex}
\eeq
plays the role of the (nonlocal) four-point function
to order $1/N$. 


In the symmetric regime we can evaluate 
(\ref{ginversexact}) 
for the configuration $G_{ab}(x,y)=G(x,y) \delta_{ab}$. In this form
the $\O (1)$ and $\O (1/N)$ contributions are explicit. We separate
$\Sigma$ in a local part and a nonlocal part,
\beq
\Sigma(x,y;G) = - i \Sigma^{\rm (local)}(x;G) \delta_{\C}^{d+1}(x-y)
+ \Sigma^{\rm (nonlocal)}(x,y;G) \, .
\label{sighominh}
\eeq
Since $\Sigma^{\rm (local)}$ corresponds to a local mass shift it is 
convenient to write 
\beq
M^2(x;G) = m^2 + \Sigma^{\rm (local)}(x;G).
\eeq 
With this notation the Schwinger--Dyson equation for the 
propagator reads 
\beq
i G^{-1}(x,y) = - \left( \square_x + M^2(x;G) \right) \delta_{\C}^{d+1}(x-y)
- i\, \Sigma^{\rm (nonlocal)}(x,y;G) \, .
\label{gevol}
\eeq 
From (\ref{sigmanext})--(\ref{ieqab}) we observe that the local part 
receives LO and NLO contributions, 
\beq
M^2(x;G) =  m^2 + \lambda \frac{N+2}{6 N}\, G(x,x)  \, ,
\label{homsignlo}
\eeq
while the
inhomogeneous part of the self energy is nonvanishing 
only at NLO,
\beq
\Sigma^{\rm (nonlocal)}(x,y;G) = - \frac{\lambda}{3 N}\, G(x,y)
I(x,y) \, ,
\label{siginh}
\eeq
where the resummed ``chain'' of bubble graphs (\ref{ieqab}) reads
\beq
I(x,y) = \frac{\lambda}{6}\, G(x,y) G(x,y)
 -\, i\, \frac{\lambda}{6} \int_{\C} {d}^{d+1}z\,
I(x,z) G(z,y) G(z,y)  \, .
\nonumber\\
\label{ieq}
\eeq   
It is important to note that for the LO, or $N \to \infty$, 
approximation only a local mass shift is induced by the interactions. 
The NLO contribution is crucial in order to include direct scattering.

\subsection{Initial-value time evolution}
\label{ssInitialvalue}

The form of the equation of motion (\ref{gevol}) is suitable for
a boundary value problem and can be used, in particular, to discuss
the propagator in thermal equilibrium.  
However, nonequilibrium time evolution is an initial value problem. 
Here a suitable form of the equation of motion can be obtained by
multiplying (\ref{gevol}) by $G$ from the right and by integration.
With $\int_{\C} {d}^{d+1}z\, G^{-1}(x,z)G(z,y)
= \delta_{\C}^{d+1}(x-y)$ one finds
\beq
\left( \square_x + M^2(x;G) \right) G(x,y) 
+\, i \int_{\C}\! {d}^{d+1}z\, \Sigma^{\rm (nonlocal)}(x,z;G) G(z,y)
= - i \delta_{\C}^{d+1}(x-y) 
\eeq
This represents a second order differential equation for 
$G$ which may be used to study the time evolution for given 
initial conditions (\ref{inG}). Very useful 
further simplification arises if we decompose 
\bea
G(x,y)&=&G_>(x,y) \Theta_{\C}(x^0-y^0)+
G_<(x,y) \Theta_{\C}(y^0-x^0) \, ,
\nonumber\\[0.2cm]
\Sigma^{\rm (nonlocal)}(x,y)&=&\Sigma_>(x,y) \Theta_{\C}(x^0-y^0)+
\Sigma_<(x,y) \Theta_{\C}(y^0-x^0) \label{sigbs} 
\eea
and observe
\beq
\square_x G(x,y)  
= \Theta_{\C}(x^0-y^0) \square_x G_>(x,y) 
+ \Theta_{\C}(y^0-x^0) \square_x G_<(x,y) 
- i \delta_{\C}^{d+1}(x-y) \, . \nonumber
\label{thetafunctions}
\eeq
Note that the equal-time commutation relation for the 
fields indeed imply
\beq
\frac{d}{d x^0}\Big( 
G_>(x,y)-G_<(x,y)
\Big)_{| x^0=y^0 }= - i \delta^d({\bf x} - {\bf y})
\label{etcr} \, .
\eeq
For a real, scalar field theory ${G_{>}}^{\!\!\!\!\! *}\,(x,y)=
G_<(x,y)= G_>(y,x)$
and all equations can be expressed in terms of one function ($G_>$) 
only. The self energy has the same properties    
${\Sigma_{>}}^{\!\!\!\!\! *}\, (x,y)=\Sigma_<(x,y)= \Sigma_>(y,x)$, 
which can
be verified from (\ref{siginh}) and is valid to all orders in $1/N$. 
The \mbox{$\Theta$--functions} in (\ref{thetafunctions}) are defined
along the time contour $C$ and can be evaluated 
\bea \lefteqn{
\int_{\C}\! {d}^{d+1}z \Sigma^{\rm (nonlocal)}(x,z;G) G(z,y)
= \int {d}{\bf z} \left\{
  \int\limits_0^{y^0} dz^0\, \Sigma_>(x,z;G) G_>(y,z) \right. }
\nonumber\\
&& \left. + \int\limits_{y^0}^{x^0} dz^0\, \Sigma_>(x,z;G) G_>(z,y)
- \int\limits_0^{x^0} dz^0\, \Sigma_>(z,x;G) G_>(z,y) \right\} \, .
\nonumber
\eea
This yields the evolution equation    
\bea \lefteqn{
\left( \square_x + M^2(x;G) \right) G_>(x,y) = 
2 \int {d}{\bf z} \Bigg\{ }
\label{gbevol}\\
&& 
\int\limits_0^{x^0} dz^0\, 
{\rm Im}\left[\Sigma_>(x,z;G)\right] G_>(z,y)
- \int\limits_0^{y^0} dz^0\, 
\Sigma_>(x,z;G)\,\, {\rm Im}\left[G_>(z,y)\right]  \Bigg\} \, ,
\nonumber
\eea
We note that (\ref{gbevol}) represents an exact  
equation for known local part $M^2(x;G)$ and 
nonlocal part $\Sigma_>(x,y;G)$ of the self energy (\ref{sighominh}).

\subsection{Real-valued evolution equations} 
\label{ssRealevolution}

For the real scalar field theory 
there are two independent real--valued two--point functions, which 
can be associated to the expectation values of the commutator and 
the anti-commutator of two fields. We note that
the field anti-commutator is determined by the real part and 
the commutator by the imaginary part of the complex function $G_>(x,y)$. 
Following Ref.\ \cite{AB1} we define  
\bea
F(x,y)&=& \frac{1}{2} \Big(G_>(x,y)+G_{>}^*(x,y)\Big) \equiv {\rm Re}[G_>(x,y)]
\label{defrho} \, ,\\[0.2cm]
\rho(x,y)&=&i \Big(G_>(x,y)-G_{>}^*(x,y)\Big) 
\equiv -2 {\rm Im}[G_>(x,y)]
\eea
where $F$ is the ``symmetric'' propagator and $\rho$ denotes the 
spectral function, with the property $F^*(x,y)=F(x,y)=F(y,x)$ and
$\rho^*(x,y)=\rho(x,y)=-\rho(y,x)$. Similarly, we define 
\bea
\Sigma_F(x,y)&=& \frac{1}{2} \Big(\Sigma_>(x,y)+\Sigma_{>}^*(x,y)\Big) 
\equiv {\rm Re}[\Sigma_>(x,y)] \, ,\\[0.2cm]
\Sigma_{\rho}(x,y)&=&i \Big(\Sigma_>(x,y)-\Sigma_{>}^*(x,y)\Big) 
\equiv -2 {\rm Im}[\Sigma_>(x,y)] \, .
\eea  

The time evolution equations for $F$ and 
$\rho$ follow from (\ref{gbevol}) \cite{AB1}
\bea \lefteqn{
\!\!\!\!\!\!\!\!\! \left( \square_x + M^2(x;F) \right) F(x,y) = 
-\int {d}{\bf z}\, \Bigg\{ }
\label{Fevol}\\
&& 
\!\! \int\limits_0^{x^0} dz^0\, 
\Sigma_{\rho}(x,z;F,\rho)\, F(z,y) 
-\int\limits_0^{y^0} dz^0\, 
\Sigma_F(x,z;F,\rho)\, \rho(z,y) \Bigg\}  \, , \nonumber
\eea
\beq
\left(\square_x + M^2(x;F) \right) \rho(x,y) = 
-\int {d}{\bf z} 
\int\limits_{y^0}^{x^0} dz^0\, 
\Sigma_{\rho}(x,z;F,\rho)\, \rho(z,y) \, .    
\label{rhoevol}
\eeq
The form of the evolution equations for the symmetric 
propagator and the spectral function is independent of the
approximation. They are equivalent to the exact Schwinger--Dyson
equation (\ref{ginversexact}), however, in the form (\ref{Fevol})
and (\ref{rhoevol}) they are more suitable for initial time problems.  

We note that due to the canonical 
commutation relations or (\ref{etcr}) the
spectral function obeys the 
equal-time properties 
\beq
\label{eqrhoprop}
\rho(x,y)|_{x^0=y^0} = 0, \;\;\;\;
\partial_{x^0}\rho(x,y)|_{x^0=y^0} = \delta^d(\bx-\by).
\label{rhoeqtime}
\eeq
The time evolution of the
spectral function is completely determined by (\ref{rhoevol})
and (\ref{rhoeqtime}).

In the NLO approximation $M^2(x;G)$ is given by
(\ref{homsignlo}), i.e.\
\beq
M^2(x;F) =  m^2 + \lambda \frac{N+2}{6 N}\, F(x,x)  \, ,
\label{Meff}
\eeq
and from (\ref{siginh}) and 
(\ref{ieq}) one finds
\bea
\Sigma_F(x,y;F,\rho) &=&  - \frac{\lambda}{3 N}\, \Big( F(x,y) I_F(x,y) 
-\frac{1}{4} \rho(x,y) I_{\rho}(x,y) \Big)\, ,
\label{SFFR}\\[0.3cm]
\Sigma_{\rho}(x,y;F,\rho) &=&  - \frac{\lambda}{3 N}\, \Big( 
F(x,y) I_{\rho}(x,y) 
+ \rho(x,y) I_{F}(x,y) \Big)\, 
\label{SRFR}
\eea
with\footnote{Note that the resummed ``chain'' of bubbles
(\ref{ieq}) can be written as $I(x,y)=I_>(x,y) \Theta_{\C}(x^0-y^0)+
I_<(x,y) \Theta_{\C}(y^0-x^0)$ with ${I_{>}}^{\!\!\!\!\! *}\,(x,y)=
I_<(x,y)= I_>(y,x)$. With this notation 
$I_F(x,y) = \frac{1}{2} (I_>(x,y)+I_{>}^*(x,y))={\rm Re}[I_>(x,y)]$ and
$I_{\rho}(x,y)=i (I_>(x,y)-I_{>}^*(x,y))= -2 {\rm Im}[I_>(x,y)]$.} 
\bea
I_{F}(x,y) &=&  \frac{\lambda}{6}\, 
\Big( F^2(x,y) - \frac{1}{4} \rho^2(x,y) \Big)  
\\
&-&  \frac{\lambda}{6}\, \int {d}{\bf z}
\left\{ \int\limits_{0}^{x^0} dz^0\,
I_{\rho}(x,z) \Big( F^2(z,y) - \frac{1}{4}\, \rho^2(z,y) \Big) \right. 
\nonumber\\
&& \qquad \quad \left. -\, 2 \int\limits_{0}^{y^0} dz^0\,  
I_F(x,z)  F(z,y) \rho(z,y)    \right\} \, ,
\label{IFFR}\nonumber\\
[0.3cm]  
I_{\rho}(x,y) &=& \frac{\lambda}{3}\, F(x,y) \rho(x,y) 
\label{IRFR}\\
&-&  \frac{\lambda}{3}\, \int {d}{\bf z}
\int\limits_{y^0}^{x^0} dz^0\,  
I_{\rho}(x,z) F(z,y) \rho(z,y) \nonumber \, .
\eea
We emphasize that $F$, $\rho$, $\Sigma_F$ and $\Sigma_{\rho}$
are real functions, and one observes that the evolution equations 
(\ref{Fevol})-(\ref{IRFR}) are causal and time-reversal invariant.

\subsection{Comparison with thermal equilibrium}

It is instructive to consider for a moment the quantities $F$ and 
$\rho$, $\Sigma_{F}$ and $\Sigma_{\rho}$, as well as
$I_{F}$ and $I_{\rho}$ in thermal equilibrium. The
$2PI$ effective action in thermal equilibrium, $\Gamma^{\rm (eq)}$, 
is given by the same expression (\ref{2PIaction}), with (\ref{LOcont})
and (\ref{NLOcont}) in the NLO large-$N$ approximation, if the closed 
time path is replaced by an imaginary path $\C=[0,-i\beta]$. 
Here $\beta$ denotes the inverse temperature. Since in equilibrium 
these functions depend only on the relative coordinates
it is convenient to consider the Fourier transforms 
$F^{\rm (eq)}(\omega,\bp)$, $\rho^{\rm (eq)}(\omega,\bp)$, 
$\Sigma^{\rm (eq)}_F(\omega,\bp)$,
$\Sigma^{\rm (eq)}_{\rho}(\omega,\bp)$,  
$I^{\rm (eq)}_F(\omega,\bp)$ and $I^{\rm (eq)}_{\rho}(\omega,\bp)$.  
From the periodicity (``KMS'') condition for the propagator in imaginary
time, $G(x,y)|_{x^0=0}=G(x,y)|_{x^0=-i\beta}$, one infers the 
generic equilibrium relations\footnote{In our conventions
the Fourier transforms of the real-valued antisymmetric functions
$\rho(x,y)$, $\Sigma_{\rho}(x,y)$ and 
$I_{\rho}(x,y)$ are purely imaginary.} (cf.\ also Ref.\ \cite{AB1})  
\bea
F^{\rm (eq)}(\omega,\bp)&=& -i
\Big(n_{\rm B}(\omega)+\frac{1}{2} \Big) \, \rho^{\rm (eq)}(\omega,\bp) \, ,
\\
\Sigma^{\rm (eq)}_F(\omega,\bp)&=& -i
\Big(n_{\rm B}(\omega)+\frac{1}{2} \Big) 
\, \Sigma^{\rm (eq)}_{\rho}(\omega,\bp) \, ,
\\
I^{\rm (eq)}_F(\omega,\bp)&=& -i
\Big(n_{\rm B}(\omega)+\frac{1}{2} \Big) 
\, I^{\rm (eq)}_{\rho}(\omega,\bp)
\eea
with $n_{\rm B}(\omega)=(e^{\beta \omega}-1)^{-1}$. Here we have 
used that with (\ref{ieq}) and (\ref{siginh}) the functions 
$I(x,y)$ and $\Sigma^{\rm (nonlocal)}(x,y)$ 
satisfy the same periodicity condition as $G(x,y)$. 
 
While the spectral 
function $\rho^{\rm (eq)}$ encodes the information about the spectrum
of the model, one observes that the symmetric function $F^{\rm (eq)}$ 
encodes the statistical aspects in terms of the particle distribution
function $n_{\rm B}$ and similarly for the $\Sigma^{\rm (eq)}$-- and 
$I^{\rm (eq)}$--functions. 
We note that the ratio $\Sigma^{\rm (eq)}_{\rho}(\omega,\bp)/2 \omega$ 
plays in the limit of a vanishing $\omega$-dependence
the role of a decay rate for one-particle excited states with
momentum $\bp$. The function $I^{\rm (eq)}_{\rho}(\omega,\bp)$ encodes 
information about the effective vertex 
as will be discussed in Sect.\ \ref{ssEffectivecoupling}.
In the following we return to the nonequilibrium case and
study the time evolution equations of Sect.\ \ref{ssRealevolution}.

\section{Numerical implementation}
\label{Sectnumimp}

The time evolution equations (\ref{Fevol})--(\ref{IRFR}) are 
nonlinear, integro-differential equations. Though the equations 
are in general too complicated to be tackled
analytically, they can be efficiently 
implemented and solved on a computer.
Here it is important to note that all equations are explicit
in time, i.e.\ all quantities at some later time $t_f$ can be obtained
by integration over the explicitly known functions for times $t<t_f$
with given initial conditions.
This aspect is discussed in more detail below.
In this respect, solving the initial time evolution
equations for nonequilibrium problems is simpler than solving the
corresponding gap equation of the form (\ref{gevol}) self-consistently, 
which is typically employed for the study of thermal equilibrium. 
For simplicity we consider spatially homogeneous fields where
\beq 
F(x,y)=F(x^0,y^0; \bx - \by) = \int \frac{{d} \bp}{(2 \pi)^d}\,
e^{i\bp(\bx-\by)} F(x^0,y^0; \bp)
\label{FTcont}
\eeq
and similarly for $\rho(x,y)=\rho(x^0,y^0; \bx - \by)$.
Spatially inhomogeneous fields pose no additional complication
but are computationally more expensive.

We employ a time discretization $x^0=n a_t$, $y^0=m a_t$ with stepsize 
$a_t$, $F(x^0,y^0;\bp) = F(n,m;\bp)$ and  
\bea
 \partial_{x^0}^2 F(x^0,y^0) &\mapsto& \frac{1}{a_t^2}
\Big( F(n+1,m) + F(n-1,m) - 2 F(n,m) \Big),
\label{deriv} \\[0.2cm]
\int\limits_0^{x^0} dt F(t,y^0) &\mapsto&
a_t \Big(F(0,m)/2 + \sum\limits_{l=1}^{n-1} F(l,m) + F(n,m)/2
\Big) \label{integral}  .
\eea 
The above simple discretization leads already 
to stable numerics for small enough stepsize $a_t$ (cf.\ also
Refs.\ \cite{Berges:2000ur,AB1}), but the convergence properties
may be easily improved with more sophisticated standard estimators.

The discretized equations for the time evolution of the matrices $F(n,m)$ and
$\rho(n,m)$ advance time stepwise in the ``$n$-direction'' for fixed $m$. 
As for the continuum the propagators obey 
the symmetry properties $F(n,m)=F(m,n)$ and $\rho(n,m)=-\rho(m,n)$.
Consequently, only ``half'' of the $(n,m)$--matrices have to be computed
and $\rho(n,n) \equiv 0$. Similarly, since $I_{\rho}$ is
antisymmetric in time $I_{\rho}(n,n)$ vanishes identically.
It is also useful to note that  
$\rho(n+1,n) \equiv a_t$ 
is fixed by commutation relations (cf.\ Eq.\ (\ref{rhoeqtime})). 
As initial conditions one has to specify $F(0,0;\bp)$, $F(1,0;\bp)$
and $F(1,1;\bp)$, while $\rho(0,0;\bp) = \rho(1,1;\bp) \equiv 0$ and 
$\rho(1,0;\bp) \equiv a_t$ are fixed.

The time discretized versions of the evolution equations (\ref{Fevol}) and
(\ref{rhoevol}), 
\bea \lefteqn{
F(n+1,m;\bp) = 2 F(n,m;\bp) - F(n-1,m;\bp) } \nonumber\\[0.1cm]
&& - a_t^2 \left\{ \bp^2 + m^2 + \lambda\, \frac{N+2}{6 N}
\int\! \frac{{d}^d \bk}{(2 \pi)^d}\, F(n,n;\bk)
\right\} F(n,m;\bp) 
\nonumber\\
&& - a_t^3\, \Bigg\{
\Sigma_{\rho}(n,0;\bp)\, F(0,m;\bp)/2 - \Sigma_F(n,0;\bp)\, \rho(0,m;\bp)/2
\\
&& \qquad +\sum\limits_{l=1}^{m-1} \Big(
\Sigma_{\rho}(n,l;\bp)\, F(l,m;\bp) - \Sigma_F(n,l;\bp)\, \rho(l,m;\bp)
\Big) \nonumber\\
&& \qquad +\sum\limits_{l=m}^{n-1}\, 
\Sigma_{\rho}(n,l;\bp)\, F(l,m;\bp) \Bigg\}\, , \nonumber 
\eea 
$(n \ge m)$ \footnote{For the discretization of the time integrals it is 
useful to distinguish the cases $n \ge m$ and $n \le m$. We compute the 
entries $F(n+1,m)$ from the discretized
equations for $n \ge m$ except for $n+1=m$ where we have to use 
the equations for $n \le m$.} and similarly for $\rho$ are explicit: 
Starting with $n=1$, for the time step $n+1$ one computes successively
all entries with $m=0,\ldots,n+1$ 
from known functions at earlier times. 
At first sight this property
is less obvious for the non-derivative expressions (\ref{IFFR}) for $I_F$
and (\ref{IRFR}) for $I_{\rho}$ whose form is reminiscent
of a gap equation. However, the discretized 
equation for $I_{\rho}$,
\bea \lefteqn{
I_{\rho}(n,m;\bq)=\frac{\lambda}{3} \int\! \frac{{d}^d \bk}{(2 \pi)^d}
\Bigg\{ F(n,m;\bq-\bk) \rho(n,m;\bk) } 
\nonumber\\
&& - a_t \sum\limits_{l=m+1}^{n-1}
I_{\rho}(n,l;\bq) F(l,m;\bq-\bk) \rho(l,m;\bk)  \Bigg\} \, ,
\eea
shows that all expressions for $I_{\rho}(n,m)$ are explicit: 
Starting with $m=n$ where $I_{\rho}$ vanishes one should lower 
$m=n,\ldots,0$ successively. For $m=n-1$ one obtains an explicit 
expression in terms of $F(n,m)$ and $\rho(n,m)$ known from the
previous time step in $n$. For $m=n-2$ the
RHS then depends on the known function $I_{\rho}(n,n-1)$  
and so on. Similarly, for given $I_{\rho}(n,m)$
it is easy to convince oneself that the discretized
Eq.\ (\ref{IRFR}) specifies $I_F(n,m)$ completely
in terms of $I_F(n,0),\ldots,I_F(n,m-1)$, which constitutes
an explicit set of equations by increasing $m$ successively.

It is crucial for an efficient numerical 
implementation that each step forward in time does not involve 
the solution of a self consistent or gap equation. This is 
manifest in the above discretization. The main
numerical limitation of the approach is set by the time
integrals (``memory integrals'') 
which grow with time and therefore slow down the 
numerical evaluation. Typically, the influence of early times
on the late time behavior is suppressed and may be neglected 
numerically in a controlled way \cite{AB2}.

\subsection{Continuum and infinite volume limit} 

We use a standard lattice discretization for a
spatial volume $V$ with periodic boundary conditions. 
For a spatial volume $V=(N_s a_s)^d$ with lattice spacing $a_s$ and periodic 
boundary conditions momenta are given by
\beq
\bp^2 \,\mapsto\, \sum\limits_{i=1}^d \frac{4}{a_s^2} \sin^2 
\left(\frac{a_s p_i}{2} \right) 
\, ,
\qquad p_i=\frac{2 \pi n_i}{N_s a_s}
\eeq
where $n_i = 0,\ldots, N_s-1$. The lattice introduces a 
momentum cutoff $\pi / a_s$. In order to obtain the theory in the
continuum the lattice spacing has to be sent to zero and renormalization
is necessary.   
In the following we consider numerical results in $1+1$ dimensions.
In this case the employed renormalization condition 
for the mass term, $M^2(x)|_{x^0=0}=M_{\rm INIT}^{2}$, 
as described by (\ref{Meff}) is sufficient 
to obtain finite, physical results. 

\begin{figure}[t]
\begin{center}
\epsfig{file=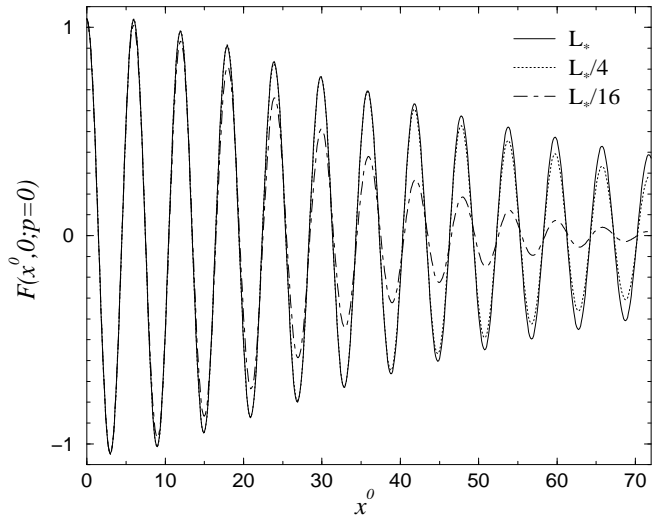,width=8.5cm,height=7.cm,angle=0}
\end{center}
\vspace*{-1.cm}
\caption{Volume dependence of the effective damping 
of oscillations,
shown for the zero mode of the unequal-time 
function $F(x^0,0;p)$ in the NLO approximation for
three different volumes. Here $L_*=256\times 0.4/M_{\rm INIT}$
denotes the ``infinite'' volume limit. The initial conditions are the same as
for \mbox{Fig.\ \ref{FigFuneq}} and the evolution is
discussed in detail below.} 
\label{Figvol}
\end{figure}
In order to study the infinite volume limit one has to remove finite 
size effects. Here this is done by increasing the volume until
convergence of the results is observed. 
\mbox{Fig.\ \ref{Figvol}} shows a typical time evolution
of the unequal-time correlation zero mode $F(x^0,0;p=0)$ for
three different volumes $L_*=N_s a_s=256\times 0.4/M_{\rm INIT}$,
$\, L_*/4\,$ and $\,L_*/16$. The initial conditions are the same as
for \mbox{Fig.\ \ref{FigFuneq}} described in detail below. One 
observes that the damping tends to be stronger for smaller volumes. We
find that increasing the volume beyond $L_*$ does not change
the results such that one could distinguish the curves in the
plot ($< 1 \%$). Depending on the initial condition and
the considered quantity the employed volumes can, however, be as large as 
$L=1600\times 0.4/M_{\rm INIT}$ as for the LO evolution shown in
{\mbox{Fig.\ \ref{FigLOM}},
or as small as $L=32\times 0.2/M_{\rm INIT}$ employed in
the second frame of 
\mbox{Fig.\ \ref{Figlambda}} showing the large-time behavior of the
effective four-point function. (For the first frame of 
\mbox{Fig.\ \ref{Figlambda}},
showing the coupling at early times, we use $L=256 \times 0.2/M_{\rm INIT}$
and one observes a small volume dependence for the 
large-time behavior of this quantity.) 
We vary our  
lattice spacings in the range 
\mbox{$a_s=(0.2\,$---$\,0.4)/M_{\rm INIT}$} and 
\mbox{$a_t/a_s=0.05\,$---$\,0.2\,$}.\footnote{    
For time evolution problems
the volume which is necessary to reach the infinite volume limit 
to a given accuracy can depend on the time scale. 
This is, in particular, due to the fact that finite systems 
can show characteristic
recurrence times after which an initial effective
damping of oscillations can be reversed. The observed damping can 
be viewed as the result of a superposition 
of oscillatory functions with differing phases or frequencies.  
The recurrence time is given by the time after which 
the phase information contained in the initial oscillations 
is recovered. Then the damping starts again
until twice the recurrence time is reached and so on.
In 1+1 dimensions we explicitly verify that in the LO 
approximation ($N \to \infty$)
the observed recurrence times for the equal-time correlation 
$F(x,x)$ scales with 
the volume $L=N_s a_s$ or the number of degrees of freedom to infinity. 
We emphasize that the phenomenon of complete recurrences, 
repeating the full initial oscillation pattern after some characteristic time,
is not observed once scattering is taken into account. 
Periodic recurrences can occur with smaller amplitudes as time proceeds and
are effectively suppressed in the large-time
limit even for small systems. As a consequence, in the NLO approximation
we find that convergence of results to a given accuracy  
can be obtained for sufficiently large, fixed volumes independently
of the time.}

\section{Nonequilibrium dynamics: damping, drifting and thermalization}
\label{Sectnonequilibriumdynamics}

The spatially homogeneous solution of (\ref{Fevol}) or (\ref{rhoevol}) 
in the limit of a free field theory ($\lambda=0$) describes modes which 
oscillate with frequency $\sqrt{{\bf p}^2+m^2}/2 \pi$ for
unequal-time functions $F(x^0,0;\bp)$ and 
$\rho(x^0,0;\bp)$, unless they are not identically zero. 
Equal-time correlation modes $F(x^0,x^0;\bp)$ oscillate either 
with twice that frequency or they are constant in time\footnote{
Note that for equal times $\rho(x^0,x^0;\bp)\equiv 0$ 
according to Eq.\ (\ref{rhoeqtime}).}. 
The latter corresponds to
solutions which are translationally invariant in time, i.e.\
$F(x^0,y^0;\bp)=F(x^0-y^0;\bp)$ and 
$\rho(x^0,y^0;\bp)=\rho(x^0-y^0;\bp)$. 

Time translation invariant solutions play an important
role for the dynamics of nonequilibrium field theory. In
the presence of interactions
we will observe below that for a large variety of initial conditions 
all modes approach time translation invariant solutions 
at asymptotically large times. This qualitative property of the asymptotic
solution can be observed both for the LO and the NLO approximation.
However, at LO the large-time behavior depends explicitly on the 
(conserved) initial particle number distribution and thermalization 
cannot be observed as is shown below. In contrast, in the NLO 
approximation scattering is
taken into account and the nonequilibrium late time result turns out
to be insensitive to the detailed initial conditions, approaching 
a thermal equilibrium distribution which is determined by the (conserved)
initial energy density. 
Following Ref.\ \cite{Wetterich:1997rp} we will call in the following time
translation invariant solutions of the evolution equations fixed
point solutions.  

We stress that during the nonequilibrium time evolution 
there is no loss of details about the
initial conditions in any strict sense. The evolution equations 
are time reflection invariant and at any time the evolution can 
be reversed and the initial conditions recovered\footnote{
In this paper we consider closed systems without coupling to an
external heat bath or external fields, which could provide 
sources or sinks of energy.}. In particular,
starting away from a time translation invariant solution
no fixed point solution can ever be reached during the 
nonequilibrium evolution 
(cf.\ also the discussion in \mbox{Sect.\ref{ssQuench}}).
However, fixed point solutions 
can be approached arbitrarily closely for sufficiently 
large times as is discussed below.

In the following we present our numerical results and provide
a detailed explanation of the summary in Sect.\ \ref{soverview}.  
To make contact with earlier work 
\cite{LOapp} we begin with a short discussion of the dynamics for the 
LO approximation 
which can be characterized by a fixed point solution \cite{Test}. 
The LO fixed point is unstable once scattering is taken into account
and we present the comparison with the NLO approximation afterwards.

\subsection{The LO fixed point}
\label{ssLOfixedpoint}

The LO contribution to the  
effective action (\ref{LOcont}) adds a time dependent mass shift 
to the free field evolution equation. The resulting effective mass term 
$M^2(x)$, given by (\ref{Meff}) for $N \to \infty$, is the 
same for all Fourier modes and each mode propagates ``collisionlessly''.
The evolution equation (\ref{Fevol}) for $F(x,y)$ 
becomes nonlinear. However, it remains local in time since the 
``scattering'' contributions 
on the RHS are absent in the LO approximation. In this case
one observes that the evolution of $F$ decouples from $\rho$.
In LO the spectral function does not influence the time evolution 
of the symmetric propagator similar to the free field theory limit.
\begin{figure}[t]
\begin{center}
\epsfig{file=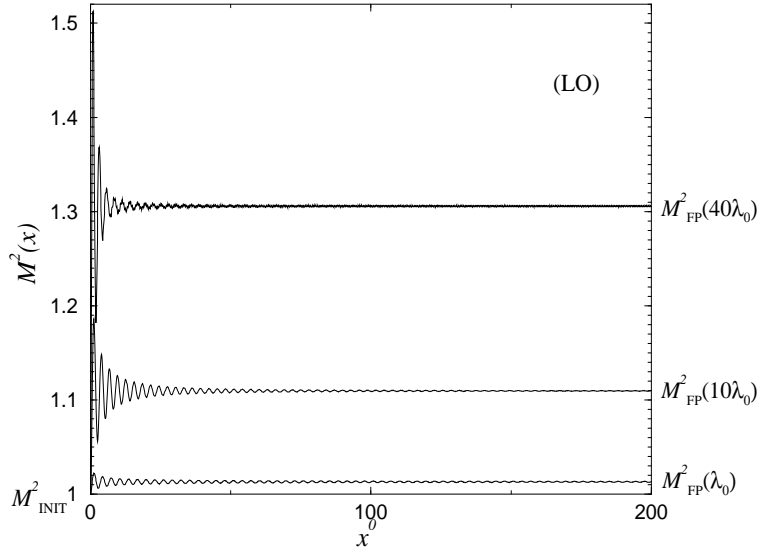,width=10.cm,height=7.5cm,angle=0}
\end{center}
\vspace*{-1.1cm}
\caption{
Shown is the mass term $M^2(x)$ in the LO approximation for
three different couplings $\lambda=\lambda_0,10\lambda_0,40\lambda_0$
in units of appropriate powers of $M^2_{\rm INIT}$. 
The asymptotic constant $M^2_{\rm FP}$ depends on the coupling
and the initial particle number distribution $n_0$. The value 
$M^2_{\rm FP}$ is well described by the LO gap
equation (\ref{LOgapequ}). The nonthermal LO fixed 
point becomes unstable once scattering is taken into account 
in the NLO approximation.} 
\label{FigLOM}
\end{figure} 
   
\subsubsection{\it ``Quench''}
\label{ssQuench}

In Fig.\ \ref{FigLOM} we plot the evolution of 
$M^2(x)$ in the LO approximation as a function of time $x^0$.
The time evolution follows a ``quench'' described by an instant drop in the
effective mass term from $M_0^2= 2 M_{\rm INIT}^2$ to 
$M_{\rm INIT}^2$. (Cf.\ the discussion in \mbox{Sect.\ \ref{soverview}}.)
The initial equilibrium high temperature ($T_0=2 M_{\rm INIT}$) 
particle number distribution is $n_0(p)=1/(\exp[\sqrt{p^2+M_0^2}/T_0]-1)$. 
The sudden change in the effective mass
term drives the system out of equilibrium and one can study its
relaxation. We present $M^2(x)$ for three different couplings 
$\lambda=\lambda_0=0.5 \, M_{\rm INIT}^2$ (bottom), $10\lambda_0$ (middle)
and $40\lambda_0$ (top).   
For the plot all quantities are given in units of appropriate powers 
of the initial time mass $M_{\rm INIT}=M(x)|_{x^0=0}$.
One observes that $M^2(x)$ shoots up from $M_{\rm INIT}^2$ in response 
to the ``quench''. The initial oscillations with frequency
$M_{\rm INIT}/\pi$ are damped and at large times $M^2(x)$ approaches 
a constant $M_{\rm FP}^2$ depending on $n_0$ and $\lambda$. 

If a strictly time translation invariant solution 
were reached, with $F(x,x)=F(x-x)$, then the
LO mass term would be a constant and given by 
$M^2=m^2+\frac{\lambda}{6}\, F(x-x \equiv 0)= M_{\rm INIT}^2$ 
according to Eq.\ (\ref{Meff}) for $N \to \infty$. Clearly 
for the initial conditions employed in \mbox{Fig.\ \ref{FigLOM}} 
the mass term is non-constant and the asymptotic values 
deviate from $M_{\rm INIT}^2$. We may compare 
the asymptotic dynamics with the solution of the LO equations 
for given particle number distribution $n_0(p)$ and an initial
mass term such that $M^2(x)=M_{\rm GAP}^2$ is constant. 
For a constant mass term the LO approximation 
behaves similar to a free theory. The propagator modes 
$F(x^0-y^0;p)=\left[ n_0(p)+1/2 \right]/\omega (p)\, 
\cos \left[\omega (p)\, (x^0-y^0)\right]$
oscillate with $\omega=\sqrt{p^2+M_{\rm GAP}^2}$, where the
effective mass term according to Eq.\ (\ref{Meff}) 
is given by the LO gap equation\footnote{
Here the logarithmic divergence of the one-dimensional integral 
is absorbed into the bare mass parameter $m^2$ using the same 
renormalization condition as for the dynamical evolution in
the LO approximation, i.e.\ 
$m^2+\frac{\lambda}{6} \int \frac{dp}{2 \pi}
\left( n_0(p)+\frac{1}{2} \right) (p^2+M^2_0)^{-1/2}
= M_{\rm INIT}^2$ .}
\beq
M^2_{\rm GAP}=m^2+\frac{\lambda}{6} \int \frac{dp}{2 \pi}
\left( n_0(p)+\frac{1}{2} \right) \frac{1}{\sqrt{p^2+M^2_{\rm GAP}}} \, .
\label{LOgapequ}
\eeq
The result from this gap equation is 
$M^2_{\rm GAP} = \{1.01,1.10,1.29\} M_{\rm INIT}^2$ for the three values
of $\lambda$.  For this wide range of couplings the values are in 
good numerical agreement with the corresponding dynamical large time 
results inferred from Fig.\ \ref{FigLOM} as 
$M_{\rm FP}^2 = \{1.01,1.11,1.31\}M_{\rm INIT}^2$ 
at $x^0=200/M_{\rm INIT}$. We explicitly checked that at
$x^0=400$ these values are still the same.  
 
We stress that the particle number does not change during the
LO time evolution and no approach to equilibrium is observed. 
Indeed the particle number   
\beq
n(p)+\frac{1}{2}
= \Big(F(x^0,y^0;p)\, \partial_{x^0}\partial_{y^0} F(x^0,y^0;p)
- \left(\partial_{x^0} F(x^0,y^0;p)\right)^2\Big)^{\frac{1}{2}}|_{x^0=y^0}
\label{pn}
\eeq 
is a strictly conserved quantity in the 
LO approximation \cite{CHKM,LObeyond,Test}.
For the quench discussed above  
$\partial_{x^0} F(x^0,y^0;p)|_{x^0=y^0}\equiv 0$ with $n(p)=n_0(p)$
and the initial high temperature distribution remains constant.

We conclude that for the spatially homogeneous situation considered here
the LO approximation cannot describe the approach to 
thermal equilibrium. The late time behavior is characterized by the
LO fixed point which explicitly depends on the initial
particle number distribution. This nonthermal fixed point becomes unstable 
once scattering is
taken into account as is discussed in the following. Before we
address the nonequilibrium dynamics in the NLO approximation in more
detail we note from Eqs.\ (\ref{Fevol}) and (\ref{rhoevol}) that the  
LO approximation for $F(x,y)$ and $\rho(x,y)$ becomes exact\footnote{
This is due to the fact that we choose to
consider Gaussian initial conditions, where at $x^0,y^0 = 0$ irreducible
$n$-point functions vanish identically for $n>2$.} 
for $x^0,y^0 \to 0$. For very early times one therefore expects
the LO approximation to yield a quantitatively valid description.
This expectation is the basis for the many applications of the
LO approximation found in the literature so far. Within the
present framework we can test when the LO approximation breaks
down by comparing it with the NLO results. 
We will see that the LO approximation fails
to describe the main qualitative aspects of an early time
exponential damping or late time approach to thermal 
equilibrium. However, for weak enough couplings certain 
time-averaged quantities 
at early times can be well determined by the LO fixed point 
solution. The particular role of the LO fixed point solution
for time-averaged quantities at early times was pointed out 
for the classical field theory limit in Ref.\ 
\cite{Test}.

\subsection{Early-time exponential damping}
\label{ssEarlytime}

In contrast to the LO approximation at NLO in the $1/N$ expansion
the evolution equations (\ref{Fevol}) and (\ref{rhoevol}) include 
the nonlocal parts of the self energy, $\Sigma_F$ and $\Sigma_\rho$, 
which encode scattering effects. In particular, 
the equations become nonlocal in time and the evolution depends
on the time history. As pointed out above these memory effects  
can already be observed on the level of the exact equations (\ref{Fevol}) and 
(\ref{rhoevol}) for known $\Sigma_{F,\rho}$. The time structure of 
the equations reflects causality and time reflection invariance which 
is preserved in the $1/N$ expansion scheme.     
\begin{figure}[t]
\begin{center}
\epsfig{file=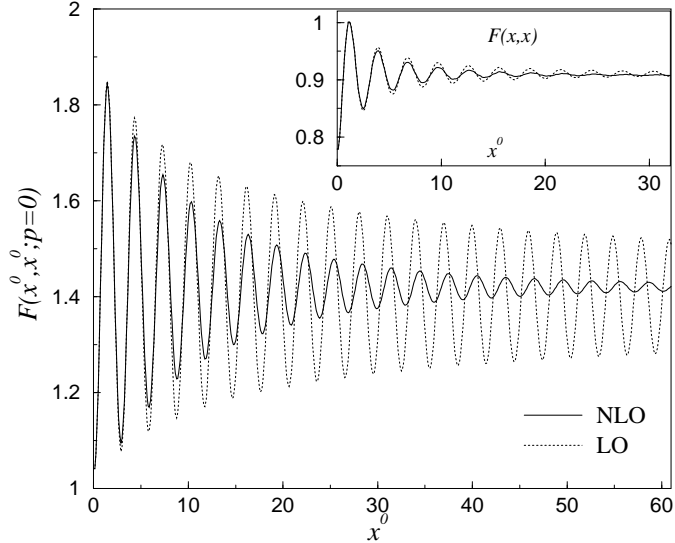,width=8.8cm,height=7.5cm,angle=0}
\end{center}
\vspace*{-1.cm}
\caption{Time dependence of the equal-time zero mode  $F(x^0,x^0;p=0)$
after a ``quench''. The employed coupling is 
$\lambda/6N = (5/6N\simeq 0.083) \, M_{\rm INIT}^2$ for $N=10$.
The inset shows $F(x,x)$ which is the 
sum over all momentum modes. (All in units of 
appropriate powers of $M_{\rm INIT}$.)
} 
\label{FigLOF}
\end{figure} 

\begin{figure}[t]
\begin{center}
\epsfig{file=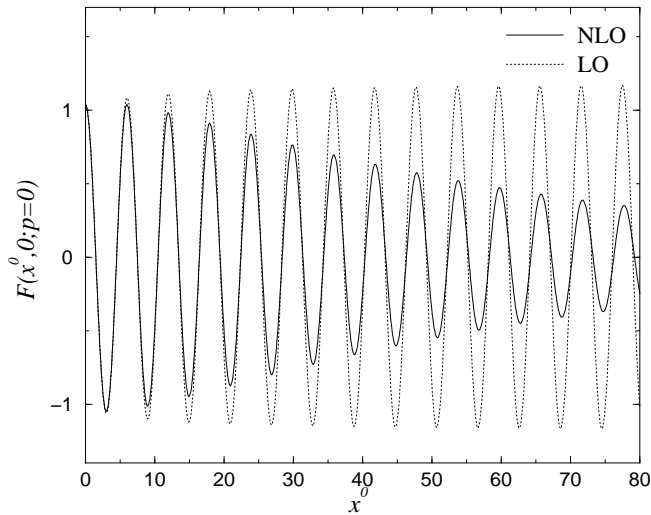,width=8.5cm,height=7.0cm,angle=0}
\end{center}
\vspace*{-1.cm}
\caption{Shown is the evolution of the unequal-time correlation 
\mbox{$F(x^0,0;p=0)$} after a ``quench''. Unequal-time 
correlation functions approach zero in the NLO approximation
and correlations with early times are effectively suppressed.
(In units of $M_{\rm INIT}$.)
} 
\label{FigFuneq}
\end{figure} 
\begin{figure}[t]
\begin{center}
\epsfig{file=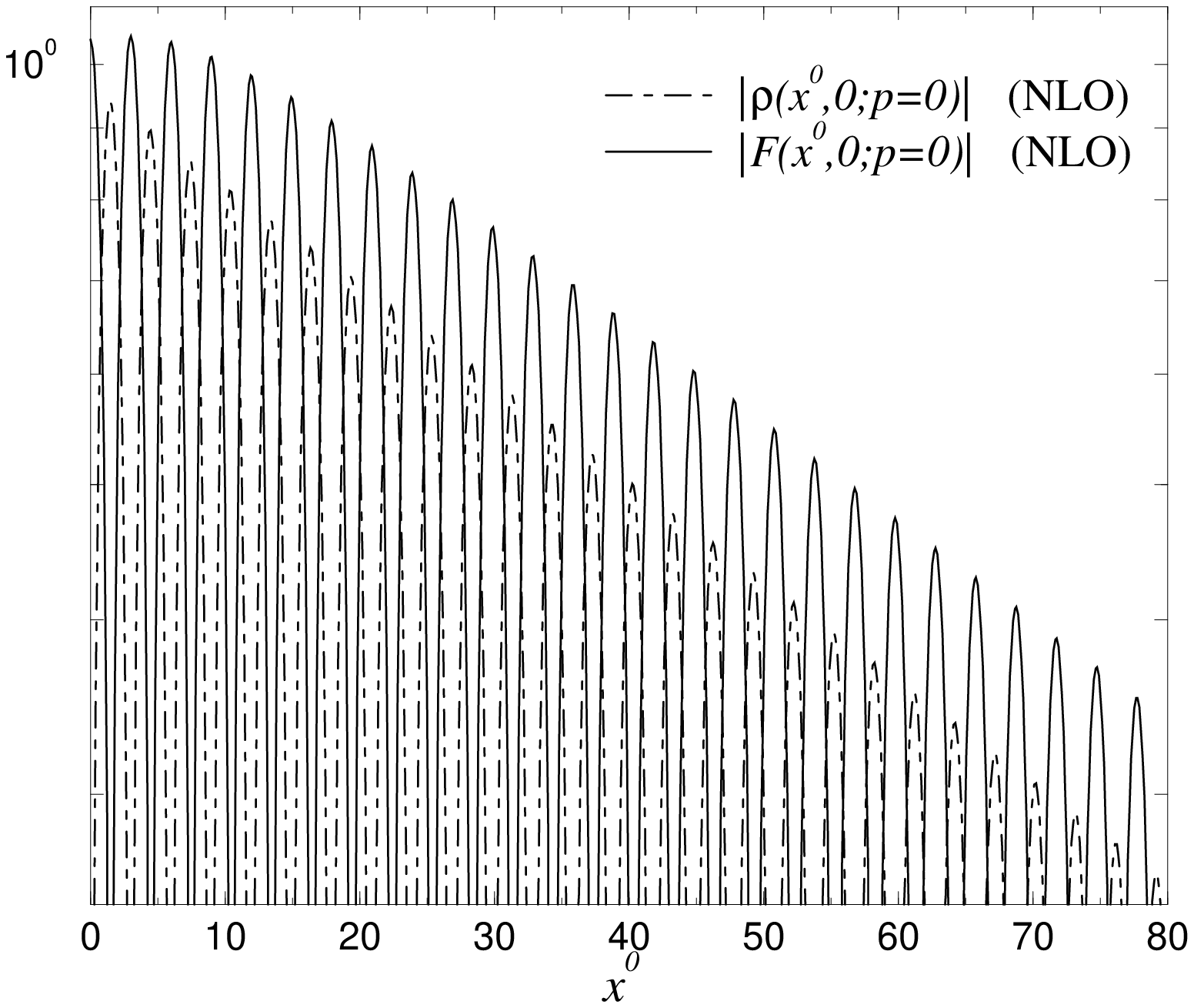,width=8.5cm,height=7.4cm,angle=0}
\end{center}
\vspace*{-1.cm}
\caption{
The correlation modes oscillations quickly approach an exponentially
damped behavior. The logarithmic plot of $|\rho(x^0,0;p=0)|$
and $|F(x^0,0;p=0)|$ as a function of time $x^0$ shows an oscillation
envelope approaching a straight
line (in units of $M_{\rm INIT}$). 
} 
\label{FigLogFrho}
\end{figure} 
In Fig.\ \ref{FigLOF} the solid line shows the behavior of the
zero momentum mode of the
equal-time correlator $F(x^0,x^0;p)$ in the NLO approximation after a
``quench''. The inset displays
the sum over all equal-time modes, 
$F(x,x)=\int \frac{d p}{2 \pi} F(x^0,x^0;p)$, and 
$M^2(x)|_{x^0=0}=M^2_{\rm INIT}$.
The coupling is $\lambda/6N = (5/6N\simeq 0.083) \, M_{\rm INIT}^2$ 
for $N=10$ and    
the initial particle number distribution $n_0(q)$ is 
the same as described in \mbox{Sect.\ \ref{ssLOfixedpoint}} for the
LO approximation. For comparison the dotted lines in 
\mbox{Fig.\ \ref{FigLOF}}
show the corresponding LO evolution.

\subsubsection{\it Damping}

As expected from the previous discussion, for very early times
the NLO behavior is well described by the 
LO approximation. However,
one observes that the damping of oscillations 
is more efficient at NLO due to scattering effects. The amplitude of the
zero mode $F(x^0,x^0;p=0)$ in \mbox{Fig.\ \ref{FigLOF}} at NLO shows a 
substantial difference to the 
LO one. Summed over all momentum modes, however, the difference 
becomes less as can be inferred from $F(x,x)$ shown in the inset of
the Figure.    
The different damping behavior becomes even more pronounced for unequal-time 
correlators. In \mbox{Fig.\ \ref{FigFuneq}} we plot $F(x^0,0;p=0)$
as a function of time. In sharp contrast to the NLO result,
where the maximum amplitude of the correlation mode is damped, 
one observes that the LO correlator is not damped at all
--- in accordance with the LO fixed point behavior described in 
the previous section. We find that the unequal-time 
two-point functions approach zero in the NLO approximation
and correlations with early times are therefore suppressed
once scattering is taken into account. We stress 
that time-reversal invariance implies that the 
oscillations can never be damped out to zero
completely during the nonequilibrium time evolution,
however, zero can be approached arbitrarily closely.

In the NLO approximation we find that 
all modes $F(x^0,y^0;p)$ and $\rho(x^0,y^0;p)$
approach an approximately exponential damping behavior for both
equal-time and unequal-time correlations. 
In Fig.\ \ref{FigLogFrho} the approach to an exponential behavior 
is demonstrated for
the unequal-time functions $|\rho(x^0,0;p=0)|$ and $|F(x^0,0;p=0)|$.
The logarithmic plot shows that after a non-exponential 
period at very early times the envelope of oscillations
can be well approximated by a straight line. From an asymptotic envelope 
fit of $F(x^0,0;p=0)$ to an exponential 
form $\sim \exp(-\gamma_0^{\rm (damp)} t)$
we obtain a damping rate $\gamma_0^{\rm (damp)}=0.016 M_{\rm INIT}$.
 
We conclude that, though strict dissipation can never
be observed in a time-reversal invariant and energy conserving
system, characteristic features of dissipative systems 
like exponential damping can be realized to very good approximation. 
The fact that correlation functions show an effective exponential 
damping once scattering is taken into account seems not
to depend on the details of the approximation. We have 
observed the same qualitative feature in a three-loop
expansion of the $2PI$ effective action \cite{Berges:2000ur,AB1}.

\subsubsection{\it Effective mass}
\label{ssEffmass}

One observes from Fig.\ \ref{FigLOF} 
that the correlation mode $F(x^0,x^0;p=0)$ shoots up 
after the ``quench'' and relaxes. The initial oscillation
frequency of the equal-time mode, $M_{\rm INIT}/\pi$, 
changes only slowly. The early-time behavior
of the NLO mass term $M^2(x)$ is rather well described by an
oscillation around the LO fixed point value. The latter
can be estimated from the solution of the LO gap equation 
(\ref{LOgapequ}), as described
in \mbox{Sect.\ \ref{ssLOfixedpoint}}, which 
yields $1.10 M_{\rm INIT}^2$. Comparing this with the
average value for $M^2(x)$ for the time interval
$0 < x^0 < 60/M_{\rm INIT}$ we find a similar value 
$\overline{M}^2=1.13 M_{\rm INIT}^2$.
 
One may ask if the agreement can be further
improved for not too late times by comparing it with an
``improved'' LO (Hartree) approximation, LO$^+$, that takes into account the
local part of the NLO self energy contribution (cf.\ Sect.\
\ref{ssSchwingerDysoneq}). In this approximation
one neglects the memory integrals on the RHS of the
NLO evolution equations (\ref{Fevol}) and (\ref{rhoevol}). 
The resulting equations are local in time and have the same structure as
the LO ones, however, with the LO and NLO contribution to 
the mass term $M^2(x)$ included as given by Eq.\ (\ref{Meff}).
The large-time limit of the mass term in the LO$^+$ approximation
is determined by the LO$^+$ fixed point solution in complete analogy
to the discussion in \mbox{Sect.\ \ref{ssLOfixedpoint}}. The
fixed point value can be estimated from the corresponding 
gap equation which yields the asymptotic LO$^+$ value
$1.11 M_{\rm INIT}^2$. The value slightly improves 
the LO result.  

\begin{figure}[t]
\begin{center}
\epsfig{file=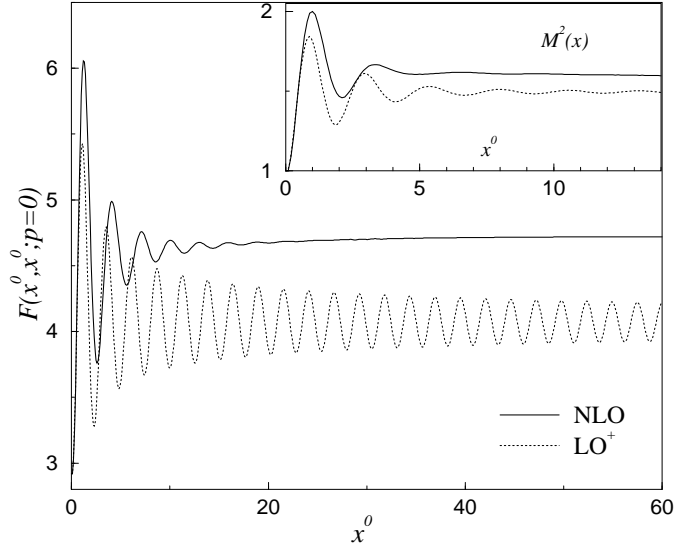,width=8.8cm,height=7.5cm,angle=0}
\end{center}
\vspace*{-0.9cm}
\caption{
Time dependence of the equal-time zero mode  $F(x^0,x^0;p=0)$
after a ``quench'', similar to Fig.\ \ref{FigLOF} but not in the 
weak coupling regime (here $2 \pi \gamma_0^{\rm (damp)}/\epsilon_0 
\sim {\cal O}(1)$). The inset shows the mass term $M^2(x)$.} 
\label{FigNLOFM}
\end{figure} 
Figs.\ \ref{FigLOF}--\ref{FigLogFrho} gave an example
for the nonequilibrium evolution after a ``quench'' for
weak effective coupling 
where the damping rate is much smaller than 
the frequency of oscillations, i.e.\ 
$1 \gg \gamma_0^{\rm (damp)}\,/(\sqrt{\overline{M}^2}/2\pi)=0.09$ for
the employed coupling $\lambda/6N = 0.083 \, M_{\rm INIT}^2$. 
In the following we consider a ``quench'' with a larger drop in the
effective mass term $M_0^2/M_{\rm INIT}^2= 2.91$ and a  
stronger effective coupling 
$\lambda/6N = 0.17 \, M_{\rm INIT}^2$ for $N=4$.
The initial 
particle number distribution is $n_0(p)=1/(\exp[\sqrt{p^2+M_0^2}/T_0]-1)$
with $T_0=8.47 M_{\rm INIT}$. As a consequence of the larger
effective coupling the correlation functions exhibit stronger
damping. The damping rate obtained from
an exponential fit to the asymptotic behavior of $F(x^0,0;p=0)$ 
is found to be $\gamma_0^{\rm (damp)}=0.11 M_{\rm INIT}$. We observe 
that the oscillation
frequency of $F(x^0,0;p=0)$ quickly stabilizes around 
$1.10 M_{\rm INIT}/2\pi=0.18 M_{\rm INIT}$ which is of the 
same order than the damping rate. Correspondingly, one finds
from \mbox{Fig.\ \ref{FigNLOFM}} an equal-time zero mode 
$F(x^0,x^0;p=0)$ which is effectively damped out at NLO
after a few oscillations. The deviation from the LO$^+$
result is substantial. 
From the inset of 
\mbox{Fig.\ \ref{FigNLOFM}} one observes after a short initial
period an almost constant mass term with a value 
$M^2(x)\simeq 1.6 M^2_{\rm INIT}$. The deviation of $M^2(x)$,
which contains the sum over all equal-time modes, from the
corresponding LO$^+$ value of about $1.5 M^2_{\rm INIT}$
is less severe than for the correlation modes. 

We note that the relatively large value for $M^2(x)$ 
infered from \mbox{Fig.\ \ref{FigNLOFM}} does not characterize
well the oscillation frequencies of the correlation zero modes. 
Following Ref.\ \cite{AB1} a better characterization
for the mode frequencies can be obtained from an effective
mode energy defined as   
\beq
\epsilon_p(x^0) \equiv \left(\, \frac{\partial_{x^0}\partial_{y^0} 
F(x^0,y^0;p)}{F(x^0,y^0;p)}\right)^{\frac{1}{2}}|_{x^0=y^0}
\label{disprel}
\eeq
which coincides with the 
dispersion relation of a free theory for $\lambda \to 0$. 
After a short oscillation period we find an almost constant value 
$\epsilon_0(x^0) \simeq 1.1 M_{\rm INIT}$.
\begin{figure}[t]
\begin{center}
\epsfig{file=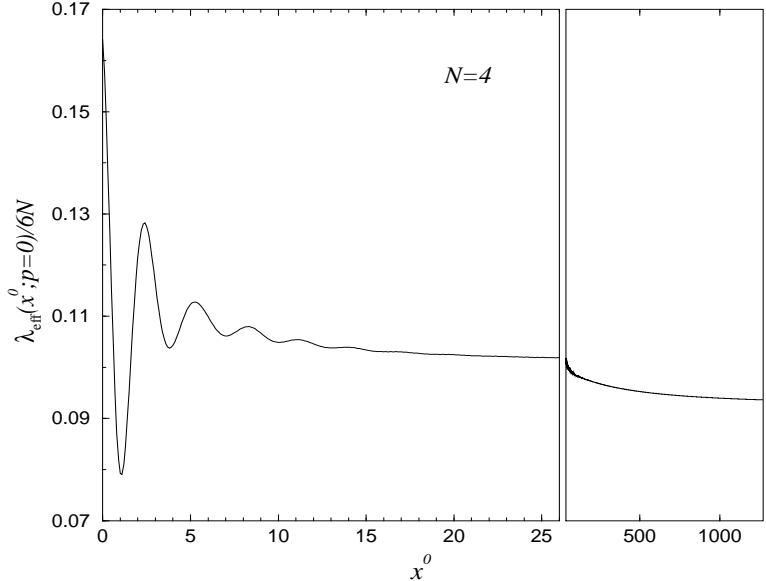,width=8.0cm,height=10.0cm,angle=-90}
\end{center}
\vspace*{-0.5cm}
\caption{Nonequilibrium evolution of the effective four-point coupling
at NLO for the initial conditions of Fig.\ \ref{FigNLOFM} 
(in units of $M_{\rm INIT}$). 
The added frame shows the large-time
behavior for $x^0 > 26/M_{\rm INIT}$ on a different scale.
} 
\label{Figlambda}
\end{figure}
One observes that $\epsilon_0/2\pi$ is in good agreement 
with the oscillation frequency for $F(x^0,0;p=0)$ given above.   
A strongly damped behavior may therefore be characterized by 
\beq
\frac{2 \pi \gamma_0^{\rm (damp)}}{\epsilon_0} \gtrsim 1 \, .
\eeq

\subsubsection{\it Effective coupling}
\label{ssEffectivecoupling}

The nonlocal NLO four-point function (\ref{fourvertex})  
can be cast into an effective local coupling by integration
over one variable. For spatially homogeneous initial conditions
we define the time-dependent effective coupling
\bea
\frac{\lambda_{\rm eff}}{6 N}(x^0;p) &=& 
\int_{\cal C} d y^0\,\, \frac{\lambda}{6 N}\, 
{\bf B}^{-1}(x^0,y^0;p) \,\, = \,\, \frac{\lambda}{6 N} \Big(
1 - i \int_{\cal C} d y^0\,\, I(x^0,y^0;p) \Big) \nonumber\\
&=& \frac{\lambda}{6 N} \Big(
1 - \int_{0}^{x^0}\!\! d^{d+1} y\,\, I_{\rho}(x^0,y^0;p) \Big) \, .
\eea
One observes that the effective coupling is determined by the
antisymmetric function $I_{\rho}(x^0,y^0;p)$ given in  
Eq.\ (\ref{IRFR}). 
\begin{figure}[t]
\begin{center}
\epsfig{file=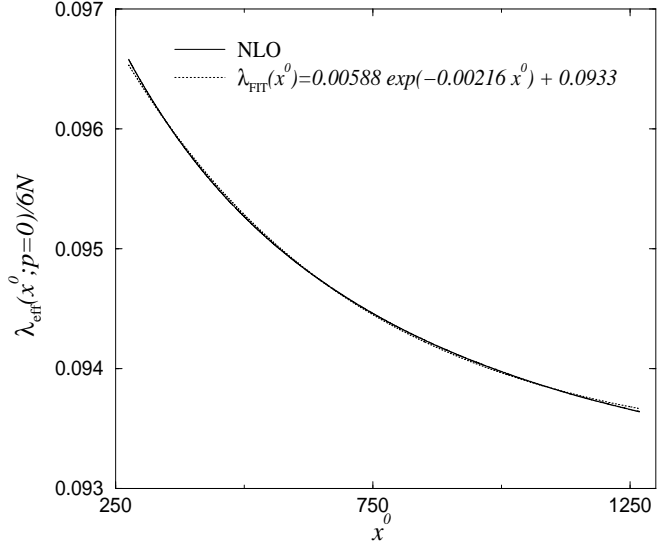,width=8.8cm,height=7.5cm,angle=0}
\end{center}
\vspace*{-0.8cm}
\caption{The same large-time behavior of the effective 
four-point coupling
at NLO as in Fig.\ \ref{Figlambda}. Shown is a comparison with the
exponential fit given in the Figure (all in units of $M_{\rm INIT}$). 
One observes for \mbox{$x^0 \gtrsim 250/M_{\rm INIT}$} 
a very good approximation by an exponential with the 
small rate \mbox{$\gamma_{\rm therm} \simeq 2.2\times 10^{-3} M_{\rm INIT}$} 
for the approach to equilibrium (cf.\ Sect.\ \ref{ssthermalization}).
} 
\label{Figlambdafit}
\end{figure}

The nonequilibrium evolution of $\lambda_{\rm eff}/{6 N}(x^0;p=0)$ is
shown in \mbox{Fig.\ \ref{Figlambda}} for the same initial
conditions as for \mbox{Fig.\ \ref{FigNLOFM}}. The effective
coupling drops in response to the ``quench'' and after a short oscillatory
period it exhibits a slowly changing behavior. We find a weak
dependence of $\lambda_{\rm eff}/{6 N}(x^0;p)$ on 
momentum. We note that for the employed Gaussian initial conditions 
the time evolution starts with the classical vertex 
$\sim \lambda$ parametrizing the action $S$, with no radiative
or thermal corrections included. At large times, when the theory
starts to thermalize, the four-point vertex approaches its 
renormalized equilibrium value as is discussed below\footnote{
In this respect the time evolution 
is reminiscent of the Wilsonian renormalization group flow
of couplings, where the flow
interpolates between the classical or microscopic
parameters and the macroscopic ones \cite{Berges:2000ew}.}.

\subsection{Non-exponential drifting at intermediate times}
\label{sDrifting}

We emphasize that the characteristic time scale for damping 
$\tau^{\rm (damp)} \sim 1/\gamma_0^{\rm (damp)}$ 
does {\it not}$\,$ correspond in general to the time scale for 
thermalization. Following a ``quench'' the initial rapid  
oscillations are effectively damped after 
$\tau^{\rm (damp)}$, however, 
the system is typically still far away from equilibrium. 
As an example, the right frame
of \mbox{Fig.\ \ref{Figlambda}} shows the asymptotic late-time behavior
of the four-point function zero mode $\lambda_{\rm eff}(x^0;p=0)/6N$
for the ``quench'' described above. Its asymptotic time dependence is 
very well approximated by an exponential behavior with rate 
$\gamma_{\rm therm}^{(\lambda)}\simeq 2.2\times 10^{-3} M_{\rm INIT}$
as is demonstrated in Fig.\ \ref{Figlambdafit}
for a wide range of times. This rate characterizes
the late-time approach to thermal equilibrium 
(cf.\ \mbox{Sect.\ \ref{ssthermalization}}). 
It is much smaller than the damping rate 
and for the employed initial condition the corresponding time scales 
differ by almost two orders of magnitude. 

We also note that the evolution of 
$\lambda_{\rm eff}$ in \mbox{Fig.\ \ref{Figlambda}}, 
after time-averaging over the oscillation time 
$\sim (\overline{\epsilon}_0)^{-1}$,
cannot be approximated by a simple exponential before
times of ${\cal O}(50/M_{\rm INIT})$. The time 
dependence is parametrically much slower than an exponential
and for $x^0 \gtrsim 1/\gamma_0^{\rm (damp)}$ better approximated 
by a power law behavior. Indeed, for a large variety of initial conditions 
we find a slow change or drifting of equal-time correlation 
modes over time
scales $\tau_{\rm drift}$ which are typically much larger than 
$\tau^{\rm (damp)}$. In the following we address
the question in more detail of what happens at intermediate times  
\mbox{$\tau^{\rm (damp)} \lesssim x^0 \lesssim \tau^{\rm (therm)}$},
between the two time scales $\tau^{\rm (damp)}$ and $\tau^{\rm (therm)}$
which are both well characterized by an exponential rate.

\subsubsection{\it ``Tsunami''}
\label{sstsun}

More than for the above ``quench'',
the characteristic drifting time scale becomes apparent 
for a ``tsunami'' 
initial condition \cite{Tsunami} (cf.\ {Sect.\ \ref{soverview}}).
In the following we consider an
initial particle number $n_0(p)$ distributed by a Gaussian 
\beq
n_0(p)={\cal A} \exp\left( 
-\frac{1}{2 \sigma^2}(|p|-p_{\rm ts})^2\right)
\label{eqtsunamidist}
\eeq
peaked around $|p|=p_{\rm ts}=5 M_{\rm INIT}$ with a width determined by
$\sigma=0.5 M_{\rm INIT}$ and amplitude ${\cal A}=10$. The initial
symmetric two-point function is 
$F(x^0,y^0;p)|_{x^0=y^0=0}=(n_0(p)+1/2)/\sqrt{p^2+M^2(0)}$ 
with $\partial_{x^0}F(x^0,y^0;p)|_{x^0=y^0=0}=0$ and 
$F(x^0,y^0;p)\, \partial_{x^0}\partial_{y^0} 
F(x^0,y^0;p)|_{x^0=y^0=0}=[n_0(p)+1/2]^2$. 
The distribution is symmetric under $p \to -p$ and reminiscent
of two colliding wave packets in one spatial dimension. The units are
given in terms of the 
renormalized initial {\it vacuum} mass (\ref{Meff}) determined 
by $M_{\rm INIT}=M(x)|_{x^0=0}$ for $n_0(p)\equiv 0$. 
Note that this is different from the initial effective
mass $M(x)|_{x^0=0}$ for the nonvanishing particle
distribution, $n_0(p) \not = 0$, as used above.
\begin{figure}[t]
\begin{center}
\epsfig{file=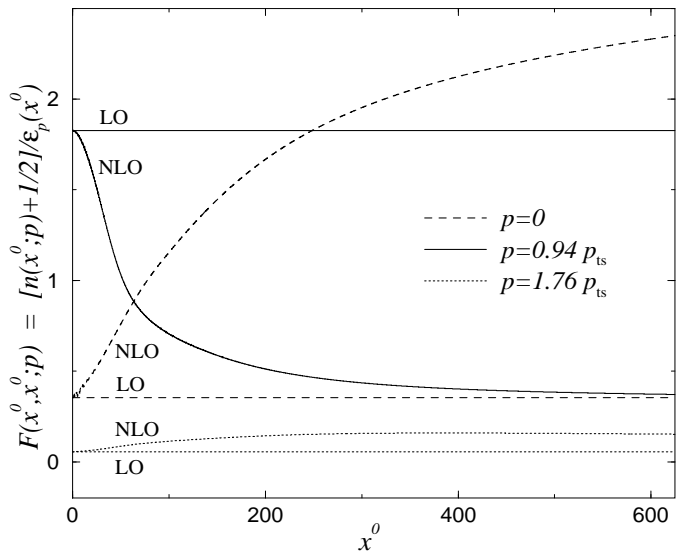,width=8.8cm,height=7.5cm,angle=0}
\end{center}
\vspace*{-1.cm}
\caption{Comparison of the LO and NLO time dependence of the 
equal-time correlation modes
$F(x^0,x^0;p)$ for the ``tsunami'' initial condition. 
The importance of scattering included in the NLO approximation
is apparent: the correlation modes $F(x^0,x^0;p)$ 
drift away from the LO result and the ``tsunami'' decays,
approaching thermal equilibrium at large times.  
} 
\label{Figpeakphi}
\end{figure} 

Similar to what is observed for the ``quench'' in 
Fig.\ \ref{FigLogFrho}, the unequal-time 
correlation modes
$F(x^0;0;p)$ and $\rho(x^0;0;p)$ for the ``tsunami'' 
in the NLO approximation
quickly approach an exponential
damping behavior. Here the damping
rate for the zero modes is $\gamma_0^{\rm (damp)}=0.022 M_{\rm INIT}$.
If we compare this with their oscillation frequency or 
$\epsilon_0$, as discussed below Eq.\ (\ref{disprel}),
one finds
\beq
2 \pi \gamma_0^{\rm (damp)}/\epsilon_0 \simeq 0.01 \, .
\label{wdt}
\eeq
The considered ``tsunami'' is clearly in the weakly damped
regime, which is due to a small effective coupling 
$\lambda/6N = 0.1 M_{\rm INIT}^2$
with $N=10$. 

\begin{figure}[t]
\begin{center}
\epsfig{file=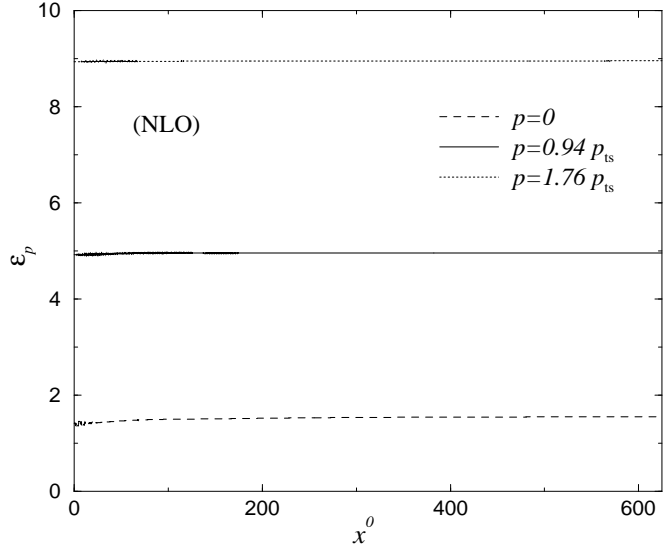,width=8.8cm,height=7.5cm,angle=0}
\end{center}
\vspace*{-1.cm}
\caption{Effective mode energy $\epsilon_p(x^0)$, defined in Eq.\
(\ref{disprel}), for the ``tsunami'' initial condition of Fig.\
\ref{Figpeakphi}. Comparison with the latter figure shows that
despite substantial changes in the effective particle
number the dispersion relation remains almost unchanged.} 
\label{Figeps}
\end{figure} 
In Fig.\ \ref{Figpeakphi} we present the time evolution of the
equal-time correlation modes $F(x^0,x^0;p)$
for different momenta. Shown are the LO and NLO results 
for zero momentum, close to the ``tsunami'' momentum
$p_{\rm ts}$ and about twice $p_{\rm ts}$. 
One observes that the LO equal-time correlations are constant
in time.
This behavior can be understood from the fact that for the 
employed ``tsunami'' initial condition 
the evolution starts at a time translation invariant fixed point 
solution of the LO equations described in \mbox{Sect.\ \ref{ssLOfixedpoint}}.

Scattering effects included in the NLO approximation drive
the system away from the LO fixed point. 
The equal-time correlation modes can be interpreted as the ratio of 
effective particle number\footnote{For time-translation
invariant two-point functions this definition corresponds to (\ref{pn}) with  
$\partial_{x^0} F(x^0-y^0;p)|_{x^0=y^0}\equiv 0$.} \cite{AB1}
\bea
n(x^0;p)+\frac{1}{2}
\equiv \Big(F(x^0,y^0;p)\, \partial_{x^0}\partial_{y^0} 
F(x^0,y^0;p) \Big)^{\frac{1}{2}}|_{x^0=y^0}
\label{eqpartnr}
\eea
and effective mode energy $\epsilon_p(x^0)$ as defined in 
\mbox{Eq.\ (\ref{disprel})}, i.e.\ 
$F(x^0,x^0;p)\equiv [n(x^0;p)+1/2]/\epsilon_p(x^0)$.  
From \mbox{Fig.\ \ref{Figpeakphi}}
one observes that the zero momentum mode (dashed line) gets more and more
populated  as time proceeds, while the high momentum 
mode (dotted line) exhibits relatively small changes.
The highly populated modes near $|p_{\rm ts}|$ become less populated
and the ``tsunami'' decays with time (solid line). 

For comparison we show in Fig.\ \ref{Figeps} the time dependence 
of $\epsilon_p(x^0)$ for the corresponding momenta. As pointed out
in \mbox{Sect.\ \ref{ssEffmass}} this quantity gives a good
description for the oscillation frequencies of the two-point function
modes. One observes from the figure
that after a short oscillatory period and a small increase of
$\epsilon_p(x^0)$ for low momenta, the effective mode energy
is almost constant in time. This is in sharp contrast to the
substantial changes in the effective particle number as can be
infered from \mbox{Fig.\ \ref{Figpeakphi}}. 
A similar observation has been pointed out in \mbox{Ref.\ \cite{AB1}}
using the three-loop approximation. 

\subsubsection{\it Power law drifting}

\begin{figure}[t]
\begin{center}
\epsfig{file=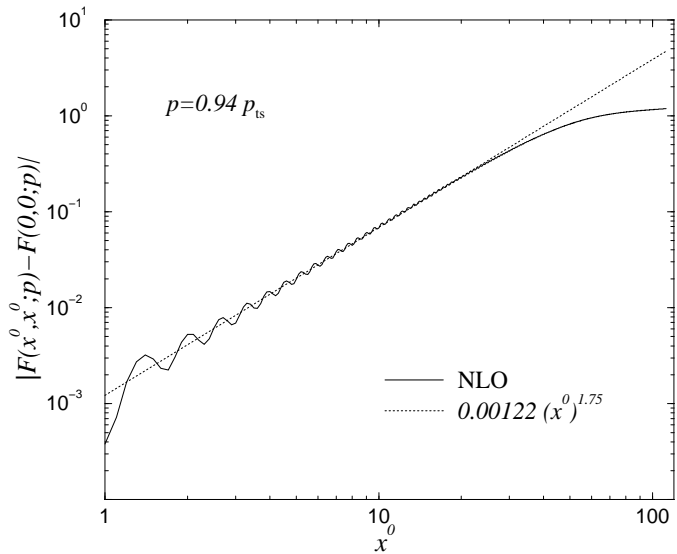,width=8.8cm,height=7.5cm,angle=0}
\end{center}
\vspace*{-1.cm}
\caption{Double logarithmic plot for the equal-time mode
$F(x^0,x^0;p)$ close to the ``tsunami'' momentum $p_{\rm ts}$.
The dotted straight line corresponds to a power law behavior
$\sim (x^0\, M_{\rm INIT})^{1.75}$. The fit describes $F(x^0,x^0;p)$ 
very well for $x^0 \lesssim 30/M_{\rm INIT}$
if time-averaged over the oscillation period. 
At later times the power law behavior 
changes to an exponential 
approach to thermal equilibrium.
} 
\label{Figptsu}
\end{figure} 
We point out that the decay of the ``tsunami'' 
modes is {\it not} governed by an exponential time dependence, here
on a time scale $\tau^{\rm (drift)} \ll \tau^{\rm (therm)}$.
The deviation of the equal-time correlation modes
from the LO fixed point is rather well approximated by a 
power law behavior and thus parametrically slower than
the exponential damping or late-time behavior. In particular,
we find that the symmetric two-point function behaves
approximately as
\beq
|F(x^0,x^0;p) - F(0,0;p)| \sim (x^0)^{{\rm (power)}_p}
\eeq
This is demonstrated 
in the double logarithmic plot of \mbox{Fig.\ \ref{Figptsu}} 
for $F(x^0,x^0;p=0.94\, p_{\rm ts})$. The NLO curve (solid line) 
oscillates around, and quickly converges to, the (dotted) 
straight line fit characterizing a power law. Time-averaged over 
the oscillation period
the evolution of the equal-time mode is well 
approximated by $\sim (x^0\, M_{\rm INIT})^{1.75}$ for 
$x^0 \lesssim 30/M_{\rm INIT}$. The details of the intermediate-time 
power law behavior of the equal-time correlation modes depend on the 
mode momenta. 

The observed behavior is reminiscent of a hydrodynamic regime
connecting the exponential early-time and late-time regimes. 
One infers from \mbox{Fig.\ \ref{Figptsu}} that
at later times a strong deviation from the straight line 
occurs. For times around $x^0 \simeq 100/M_{\rm INIT}$ the
evolution is governed by neither a simple power nor exponential 
dependence. At sufficiently large times it approaches an exponential
evolution towards thermal equilibrium (cf.\ Sect.\ \ref{ssthermalization}). 

\subsubsection{\it Strong interactions at intermediate times}
\label{ssStronginteractions}

For higher 
$n$-point functions one can observe approximate power law behavior
for certain intermediate time periods as well, however, the time evolution
before the exponential late-time behavior can be rather
complex. In \mbox{Fig.\ \ref{Figlam14}} we
show the effective four-point function $\lambda_{\rm eff}(x^0;p=0)$
for the above ``tsunami'' initial condition. We find
that before times of about \mbox{$x^0 \lesssim 300/M_{\rm INIT}$}
the drifting behavior is clearly non-exponential. For later times
the evolution of $\lambda_{\rm eff}(x^0;p)$ approaches an exponential 
time dependence with a characteristic zero mode rate 
$\gamma^{\rm (therm)}_{\lambda} \simeq 1.7\times 10^{-3} M_{\rm INIT}$. 
The thermalization time exceeds the characteristic 
damping time by more than an order of magnitude, similar to what has 
been observed above for the ``quench'' initial condition. This clear 
separation of scales happens despite the fact that the considered ``tsunami''
is weakly damped according to \mbox{Eq.\ (\ref{wdt})}.

\begin{figure}[t]
\begin{center}
\epsfig{file=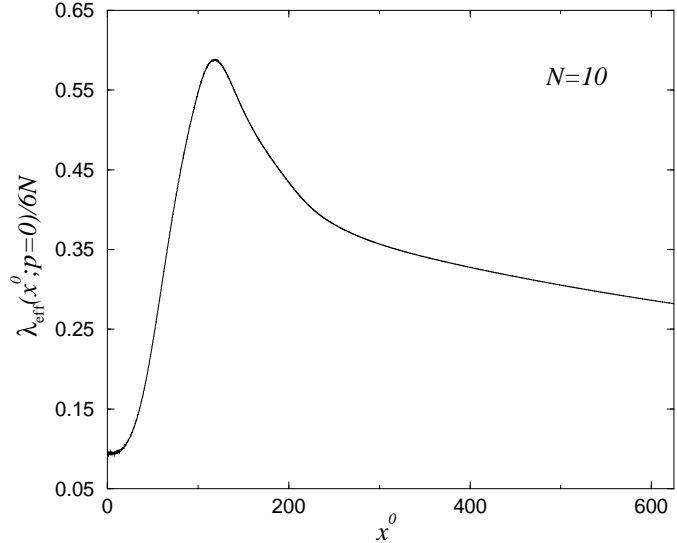,width=8.8cm,height=7.5cm,angle=0}
\end{center}
\vspace*{-1.cm}
\caption{The effective four-point function 
$\lambda_{\rm eff}(x^0;p)/6N$ at zero momentum for
the ``tsunami'' employed in \mbox{Fig.\ \ref{Figpeakphi}}.
The effective coupling for the zero momentum mode rises 
substantially at intermediate times before it approaches thermal
equilibrium. 
For $x^0 \gtrsim 300/M_{\rm INIT}$ the curve is
very well approximated by an exponential behavior with rate 
$\gamma^{\rm (therm)}_{\lambda}\simeq 1.7\times 10^{-3} M_{\rm INIT}$.
} 
\label{Figlam14}
\end{figure} 
The effective coupling can, however,
change substantially during the nonequilibrium evolution.  
From \mbox{Fig.\ \ref{Figlam14}} we observe a large increase 
of \mbox{$\lambda_{\rm eff}(x^0;p=0)$} in response to the 
``tsunami'' initial condition. The system becomes effectively 
much stronger interacting for some time before it relaxes to 
equilibrium. For the considered initial condition the phenomenon 
of a strong interaction at intermediate times is restricted to the
low momentum modes with $p \simeq 0$. For higher momenta we find that
the effective coupling changes much less between the initial and the 
late-time value. One observes that the quick
population of low momentum modes encountered in 
\mbox{Fig.\ \ref{Figpeakphi}} goes along with a large
effective coupling. Any estimate of the involved time
scale needs to take into account the strong renormalization
of the four-point function during the nonequilibrium evolution.    

We stress that a simple exponential relaxation can never exhibit
such a nontrivial behavior. 
The non-exponential drifting at intermediate
times is an important phenomenon that separates the early-time and the
late-time scales associated to exponential damping on the one hand
and thermalization on the other hand. The overall time needed to reach 
thermal equilibrium crucially depends on this phenomenon and
cannot be separated from it.

\subsection{Late-time exponential thermalization}
\label{ssthermalization}

Independent of the detailed ``quench'' or ``tsunami''
initial conditions we find for sufficiently large
times an exponential behavior to very good approximation.
This is exemplified for the effective four-point function displayed
in Figs.\ \ref{Figlambda} and \ref{Figlambdafit} after
a ``quench''. The latter
figure verifies that after a certain time the correlation function 
follows closely the exponential fit for a large time range 
$\Delta x^0 = 1000 / M_{\rm INIT}$. A similar
observation holds for the ``tsunami''
employed in \mbox{Fig.\ \ref{Figlam14}}. In particular, for both 
types of initial conditions we find that the
corresponding late-time scale exceeds the  
damping time scale by far, $\tau^{\rm (therm)} \gg \tau^{\rm (damp)}$.
In other words, we observe a very efficient damping even for
relatively weak coupling or, correspondingly, a slow thermalization.

\begin{figure}[t]
\begin{center}
\epsfig{file=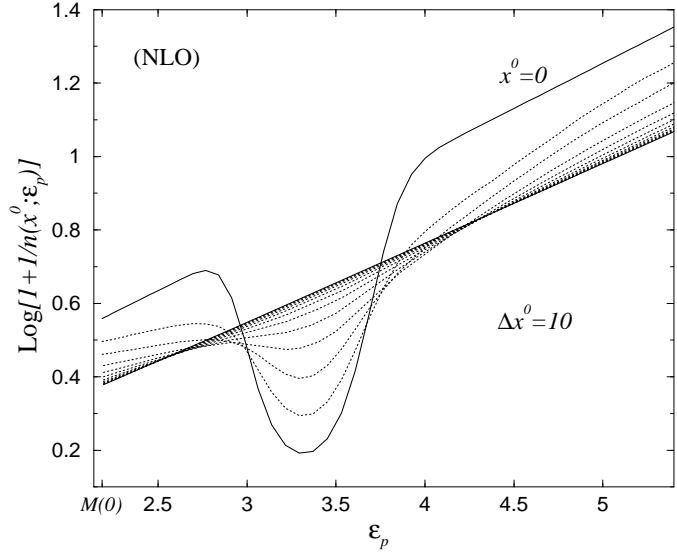,width=8.8cm,height=7.5cm,angle=0}
\end{center}
\vspace*{-1.cm}
\caption{Effective particle number distribution for a ``tsunami''
in presence of a thermal background. The solid line shows the
initial distribution which for low ($p \ll p_{\rm ts}$) and for
high ($p \gg p_{\rm ts}$) momenta follows a Bose-Einstein distribution,
i.e.\ $\ln [1+1/n(0;\epsilon_p)] \simeq \epsilon_p(0) / T_0$ 
with $T_0 = 4 M_{\rm INIT}$. At late times the nonthermal distribution
equilibrates and approaches a straight line with inverse slope
$T_{\rm EQ}=4.7\, M_{\rm INIT}\, >\, T_0$.} 
\label{Figparticlenr}
\end{figure} 
To observe thermalization on a shorter time scale we discuss
in the following a ``tsunami'' initial condition for a stronger
coupling $\lambda/6N = 0.5 \, M_{\rm INIT}^2$ 
(with $p_{\rm ts}=2.5 M_{\rm INIT}$, $\sigma=0.25 M_{\rm INIT}$
and ${\cal A}=4$ in \mbox{Eq.\ (\ref{eqtsunamidist})}).
In addition, we consider this peaked particle distribution
in presence of a thermal background with initial
``temperature'' $T_0=4\, M_{\rm INIT}$.\footnote{The initial mass term
given by Eq.\ (\ref{Meff}) is $M(0)=2.24 M_{\rm INIT}$.} The initial particle
number distribution can be infered from the solid line of 
\mbox{Fig.\ \ref{Figparticlenr}}. Shown is the combination
$\ln [1+1/n(x^0;p)]$ as a function of $\epsilon_p(x^0)$, 
where the effective particle number and mode energy 
are defined in Eqs.\ (\ref{eqpartnr}) and (\ref{disprel}) 
respectively. If $n(\epsilon_p)$ strictly follows 
a Bose-Einstein distribution with temperature $T$ then 
$\ln [1+1/n(x^0;p)] = \epsilon_p / T $. Correspondingly, from the solid
line ($x^0=0$) in \mbox{Fig.\ \ref{Figparticlenr}} one observes 
the initial ``thermal background'' as a straight line distorted
by the nonthermal ``tsunami'' peak. In the interacting theory
for $x^0 > 0$ the employed definitions for effective particle
number and mode energy should allow for a characterization of the late-time
thermal equilibrium in a similar way if a ``quasi-particle'' picture
applies (cf.\ also the discussion in \mbox{Ref.\ \cite{LOinh}}).        

Similar to the discussion of the initially weakly coupled
``tsunami'' in \mbox{Sect.\ \ref{sDrifting}}, one observes
from \mbox{Fig.\ \ref{Figparticlenr}} that the initial high 
occupation number in a small momentum range decays. More and 
more low momentum modes get excited
and the particle distribution approaches a thermal shape.
After rapid changes in $n(x^0;\epsilon_p)$ at early times
the curves representing snapshots at equidistant time steps   
$\Delta x^0=10/M_{\rm INIT}$ converge to a straight line to
high accuracy 
with inverse slope $T_{\rm EQ}=4.7\, M_{\rm INIT}\, >\, T_0$. 
The asymptotic curve therefore corresponds precisely to what one
expects of a thermally 
equilibrated ``quasi-particle'' distribution.
We find that the asymptotic late-time result for the 
distribution is determined by the initial energy
density and is insensitive to the details of the initial
conditions (``thermal fixed point''). 
The latter observation has been explicitly demonstrated before
in the three-loop approximation of the $2PI$ effective action 
in \mbox{Ref.\ \cite{Berges:2000ur}}.

We emphasize again that 
thermalization or independence of the initial 
conditions cannot be observed in a strict sense because of 
time-reflection invariance (thermal equilibrium is invariant 
under time translation and bears no information about
initial conditions). However, our results show that thermal
equilibrium can be approached very closely without deviating from it
over times of interest. Of course, the initial conditions
for which thermalization can be observed in this sense have 
to comply with standard clustering requirements or not too
large initial spread in energy\footnote{Note that the energy and its 
moments are conserved quantities during the nonequilibrium time
evolution. See also the detailed discussion in Ref.\ \cite{Aarts:2000mg} for
classical field theories.}. It is important to note that starting 
from ``first principles'' in a time-reflection invariant, energy 
conserving description thermalization can only be achieved in the 
above sense: thermal equilibrium can be approached arbitrarily closely
with time, indistinguishable for most practical purposes, however without 
reaching it on a fundamental level. This is a matter of principle
and not a question of the employed approximation.

\subsubsection{\it Inverse slope parameter}
\label{ssInverseslope}

In the following we consider
the inverse slope of the effective particle number 
distribution as plotted in \mbox{Fig.\ \ref{Figparticlenr}}. 
For low momenta 
$p \ll p_{\rm ts}$ or high momenta $p \gg p_{\rm ts}$, where the slope of
${\rm ln}(1+1/n(x^0;\epsilon_p))$ is insensitive to small 
changes in $\epsilon_p$ at fixed $x^0$,
a characterization in terms of a ``mode temperature'' $T(x^0;p)$ may be
useful --- bearing in mind that away from thermal equilibrium this 
quantity has a precise
meaning only in terms of the correlation functions employed in the
definitions (\ref{disprel}) and (\ref{eqpartnr}). 

\begin{figure}[t]
\begin{center}
\epsfig{file=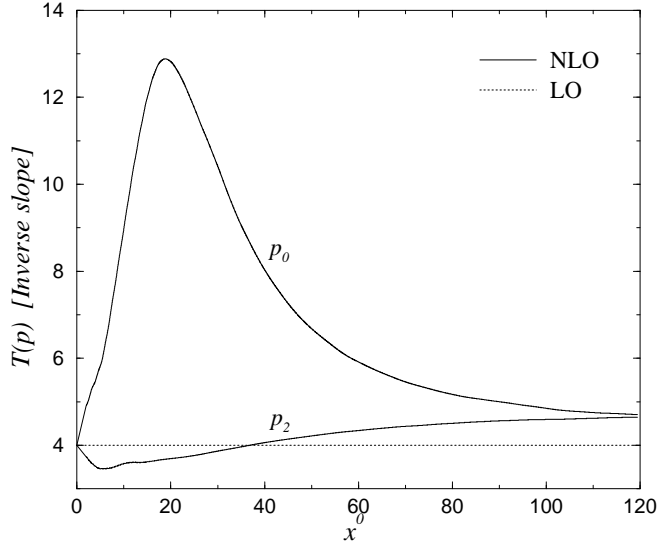,width=8.8cm,height=7.5cm,angle=0}
\end{center}
\vspace*{-1.cm}
\caption{
The time dependent inverse slope $T(x^0;p)$ of the distribution
function as shown in \mbox{Fig.\ \ref{Figparticlenr}}. The function
$T(x^0;p)$ is plotted for low momentum $p_0 \simeq 0$ and high
momentum $p_2 \simeq 2 p_{\rm ts}$, where the slope of
${\rm ln}(1+1/n(x^0;\epsilon_p))$ is insensitive to small 
changes in $\epsilon_p$ at fixed $x^0$
and a characterization in terms of a ``mode temperature'' may be
useful. At late times $T(x^0;p)$ becomes independent of momenta and
the system can be described by a global value $T_{\rm EQ}$.} 
\label{Figinvslope}
\end{figure}  
\mbox{Fig.\ \ref{Figinvslope}} shows the inverse slope $T(x^0;p)$
as a function of time for two momenta, $p_0 \simeq 0$ and 
$p_2 \simeq 2 p_{\rm ts}$ at NLO (solid line) and LO (dotted line). 
As pointed out above the particle number is conserved at LO and
the initial nonthermal distribution is ``frozen in''. Once scattering
is taken into account at NLO the distribution evolves.
One observes that for the low momentum
modes $T(x^0;p)$ shoots up in response to the ``tsunami'' 
initial condition, before dropping to the equilibrium value
$T_{\rm EQ}=4.7\, M_{\rm INIT}$. While the low momentum modes
get strongly ``overheated'' at intermediate times with respect 
to the thermal equilibrium value, one finds that the high momentum modes
exhibit a small initial drop in $T(x^0;p)$ which then smoothly 
evolves. The population of the low lying
modes involves a rapid change in the particle number distribution
in which the inverse slope parameter overshoots the late-time
value. Asymptotically the different ``mode temperatures''
converge.

It is interesting to observe that though the low momentum
modes deviate at intermediate times much stronger from 
$T_{\rm EQ}$ than the high momentum ones, there is no substantial 
difference for the equilibration time as measured e.g.\ by the
time at which they have approached $T_{\rm EQ}$ within $10$ \%.
A similar observation can be made for the ``quench''
initial conditions discussed above where we do not see that  
some modes thermalize first.

\section{Conclusion and outlook}
\label{SectCAO}

The generating functional for $2PI$ Green's functions provides 
a powerful technique for out-of-equilibrium physics. In combination
with a $1/N$ expansion it yields a controlled nonperturbative description
from ``first principles'' which is time-reversal invariant and energy 
conserving. The approach is not restricted to weak couplings
or situations close to thermal equilibrium or to an effective 
description based on a separation of scales. With the
present approach we have a quantitative method at hand which can deal
with far-from-equilibrium processes, where the latter assumptions
are typically not justified. An example is the possible 
strong renormalization of the effective four-point function at 
intermediate times for weak initial coupling 
(cf.\ Sect. \ref{ssStronginteractions}), 
which stresses the fact that a clear separation of scales can be difficult 
to achieve away from equilibrium. 

In this work we have studied two classes of initial conditions 
representing a ``quench'' and a ``tsunami'', both in the weak 
($2 \pi \gamma_0^{\rm (damp)}/\epsilon_0 \ll 1)$ and
the stronger coupling regime 
($2 \pi \gamma_0^{\rm (damp)}/\epsilon_0 \simeq {\cal O}(1))$.     
Independent of the details of the initial conditions we observe
three qualitatively different 
time regimes, as summarized in \mbox{Fig.\ \ref{Figoverview}}. 
In particular, we find 
$\tau^{\rm (damp)} \ll \tau^{\rm (therm)}$ which emphasizes that 
the damping time cannot be identified with the thermalization time. 
It is possible to relate the damping rate 
to the ``width'' of the Wigner transformed spectral function $\rho$
in the same way as discussed in \mbox{Ref.\ \cite{AB1}}, however, 
the thermalization rate cannot. Processes with nontrivial momentum 
exchange are necessary 
for thermalization and are taken into account at NLO which includes
off-shell effects\footnote{In $1+1$ dimensions on-shell two-to-two 
scattering is constrained by the energy conservation relation such
that it does not change the particle numbers for the involved momentum
modes.}. On-shell particle number 
changing diagrams as the one in \mbox{Fig.\ \ref{EYEfig}}, which appears
at next-to-next-to-leading order, can be important since they contain 
dynamics on which the bulk viscosity depends \cite{Jeon,CH}.

We emphasize that scattering effects are crucial, even for 
rather early times. None of the features
summarized in \mbox{Fig.\ \ref{Figoverview}} are captured by the 
mean field type LO approximation which neglects direct scattering; 
no exponential damping (in particular, no suppression 
of correlations $F(x^0,0;p)$ and $\rho(x^0,0;p)$ with the initial time), 
no drifting and no thermalization can be observed in that case. 
The LO approximation typically breaks down after
the first few oscillations and is approximately
valid only for $x^0 \ll \tau^{\rm (damp)}$. However, for suitably
time- or momentum-averaged quantities and not too strong couplings
we find that the LO fixed point solution provides a rather accurate 
estimate up to times $x^0 \simeq \tau^{\rm (damp)}$. The latter
observation agrees with what has been found for the corresponding
classical field theory in \mbox{Ref.\ \cite{Test}}.

We note that we are
able to observe the same qualitative properties using the
three-loop approximation of the $2PI$ effective action for
$N=1$ as employed in \mbox{Refs.\ \cite{Berges:2000ur,AB1}}. 
Despite quantitative
differences this points out that the qualitative properties
discussed here do not depend on the detailed implementation of
scattering and memory effects. In particular, this emphasizes
the importance of the setting sun type diagram of 
\mbox{Fig.\ \ref{SELFfig}}. A detailed quantitative 
comparison is deferred to a separate publication which 
also discusses universality (independence of initial 
conditions) and parametric dependences of time scales \cite{AB2}. 

There are many directions to pursue which are beyond the scope 
of the present work. With a more efficient numerics --- the computations
were done on a PC with unparallelized code --- one can directly
address similar problems in higher dimensions including the possibility
of spontaneous symmetry breaking. This would allow for a quantitative
quantum field theoretical study of the formation of disoriented chiral
condensates in the context of heavy-ion collisions or of (p)reheating
at the end of inflation in the early universe along the lines
presented here. In the present form
the time-nonlocal scattering terms are difficult to estimate
analytically. Though their numerical implementation is rather simple
the calculations are time and memory consuming. However, one may expect 
that in higher dimensions both damping and thermalization become more 
efficient which reduces the need to follow the memory kernels to very large 
times. It is important to have a 
controlled, quantitative method which can serve as the basis for 
the development of approximation schemes which may allow to find
reliable analytical solutions in limiting cases. The present
techniques present a wealth of possibilities to study the applicability
of standard phenomenological approaches. The range of validity of the 
classical Boltzmann equation or more sophisticated kinetic descriptions
can be tested as well as other typical approximation schemes based on
local equilibrium or linear response assumptions. 

We emphasize that 
the current approach is not restricted to scalar fields though
the presence of a small nonperturbative expansion parameter 
depends, of course, on the respective theory. It would be
very desirable to extend the approach to include fermions 
and gauge fields.\\

\vspace*{0.4cm}
\noindent
{\large \bf Acknowledgements}\\

\vspace*{-0.2cm}
\noindent
I thank G.\ Aarts and J.\ Cox for collaboration on related work, and
\mbox{A.\ Jakovac}, \mbox{T.\ Prokopec} and C.\ Wetterich for interesting 
discussions. {\mbox{I would} like to thank W.\ Wetzel for support with 
computers.

\end{document}